\documentclass[11pt,english]{article}
\usepackage[T1]{fontenc}
\usepackage[latin9]{inputenc}
\usepackage{amsmath}
\usepackage{esint}
\usepackage{wrapfig}

\newcommand{\fft}[2]{\frac{#1}{#2}}
\newcommand{\ft}[2]{{\textstyle\frac{#1}{#2}}}
\newcommand{\nn}{\nonumber}

\makeatletter
\usepackage{jheppub}

\title{Rigid Supersymmetric Backgrounds of 3-dimensional Newton-Cartan Supergravity}

\author[a]{Gino Knodel,}
\author[a]{Pedro Lisb\~ao,}
\author[a]{James T. Liu}

\affiliation[a]{Michigan Center for Theoretical Physics, Randall Laboratory
of Physics,\\
The University of Michigan, Ann Arbor, MI 48109--1040, USA}

\emailAdd{gknodel@umich.edu}
\emailAdd{plisbao@umich.edu}
\emailAdd{jimliu@umich.edu}

\abstract{Recently, a non-relativistic off-shell formulation of three dimensional Newton-Cartan supergravity was proposed as the $c \rightarrow \infty$ limit of three dimensional $\mathcal{N}=2$  supergravity \cite{Bergshoeff:2015uaa}. In the present paper we study supersymmetric backgrounds within this theory. Using integrability constraints for the non-relativistic Killing spinor equations, we explicitly construct all maximally supersymmetric solutions, which admit four supercharges. In addition to these solutions, there are $\frac{1}{2}$-BPS solutions with reduced supersymmetry. We give explicit examples of such backgrounds and derive necessary conditions for backgrounds preserving two supercharges. Finally, we address how supersymmetric backgrounds of $\mathcal{N}=2$ supergravity are connected to the solutions found here in the $c \rightarrow \infty$ limit.}

\makeatother

\usepackage{babel}
\usepackage{graphicx}
\begin{document}
\preprint{MCTP-15-31}
\maketitle

\section{Introduction}

The study of non-relativistic field theories and their holographic
duals has led to a renewed recent interest in Newton-Cartan gravity
\cite{Son:2013rqa,Geracie:2014nka,Gromov:2014vla,Christensen:2013lma,Christensen:2013rfa,Hartong:2014oma,Banerjee:2014nja,Hartong:2015wxa,Jensen:2014aia,Geracie:2015xfa,Fuini:2015yva}.
The latter theory is formulated as a covariant description of Newtonian
gravity, incorporating the notion of absolute time in a geometric
framework (see e.g. \cite{Misner:1974qy,Datcourt} for pedagogical
introductions). It has been argued that in the context of non-relativistic
holography, Newton-Cartan gravity is the natural geometric language
in which the bulk-boundary dictionary is to be developed. For example, the
boundary geometry of Lifshitz spacetime has been shown to be described
by a Newton-Cartan geometry with torsion \cite{Christensen:2013lma,Christensen:2013rfa}.
On the other hand, starting with a non-relativistic field theory,
Newton-Cartan gravity arises as a means of introducing (non-relativistic)
coordinate invariance: while a (relativistic) CFT may be coupled to
dynamical gravity by introducing a metric $g_{\mu\nu}$ with dynamics
governed by General Relativity, non-relativistic field theories couple
naturally to Newton-Cartan gravity, which can be formulated in terms
of two degenerate metrics, $\tau_{\mu\nu}$ and $h^{\mu\nu}$. This
insight has been used to construct effective field theories for quantum
Hall states and study universal features of the theories obtained
in this way \cite{Son:2013rqa,Geracie:2014nka,Gromov:2014vla}. 

Although the degrees of freedom in Newton-Cartan gravity differ fundamentally
from those of General Relativity, many conceptual aspects still carry
forward to the non-relativistic case. In the same way that General
Relativity can be written as a gauge theory of the Poincar\'e algebra,
Newton-Cartan gravity can be formulated as a gauge theory of the Bargmann
algebra, which is the centrally extended Galilei algebra \cite{Andringa:2010it}.
The formulation of gravity as a gauge theory has the advantage that
introducing supersymmetry to construct theories of supergravity is
relatively straightforward. In complete analogy to the case of conventional
(relativistic) supergravity, it is therefore possible to construct
supergravity theories with a non-relativistic supersymmetry group.
In three dimensions, an on-shell theory of Newton-Cartan supergravity with four
real supercharges was constructed using a vielbein approach in \cite{Andringa:2013mma}.
Moreover, by using a non-relativistic limiting procedure, the authors
of \cite{Bergshoeff:2015uaa} were able to construct an off-shell
version of the latter theory, starting from off-shell ${\cal N}=2$ 
supergravity \cite{Howe:1995zm,Achucarro:1989gm}. These recent developments allow us to ask many of the
interesting questions that arise within the context of supersymmetry
and supergravity, applied to a non-relativistic context.

The main motivation for this paper is the prospect of using Newton-Cartan
supergravity to elucidate some open questions in non-relativistic
gauge/gravity dualities. In the standard case of relativistic AdS/CFT,
much recent progress in understanding various dualities has been made
by using supersymmetric localization. This technique
allows one to calculate observables in supersymmetric theories exactly,
without having to resort to perturbation theory (for a review of recent
progress see \cite{Teschner:2014oja} and references therein). The new results obtained this way can be used to provide precision
tests of AdS/CFT: for example, the free energy of $\mathcal{N=}2^{\star}$
theories on $S^{4}$, calculated via localization \cite{Pestun:2007rz}
matches the result that one obtains from a holographic calculation
\cite{Bobev:2013cja}. Given that in the context of non-relativistic
holography, a microscopic description in terms of branes is not always
available to motivate the duality between non-relativistic field theories and gravity, 
it would be desirable to find similar precision tests for
non-relativistic AdS/CFT.

\emergencystretch=3em
Holographic results for observables in non-relativistic geometries
such as Lifshitz and Schr{\"o}dinger spacetimes are plentiful.
However, on the field theory side exact results are naturally difficult
to obtain, due to the strongly coupled nature of the theories involved.
Given the success of studying supersymmetric
theories in the relativistic case, in particular using localization, 
one concrete open question is: is there a non-relativistic analog of supersymmetric
localization? To answer this question, it is first necessary to understand
and further explore the notion of ``non-relativistic supersymmetry''
itself. While specific examples of non-relativistic supersymmetric
field theories have been constructed previously \cite{Leblanc:1992wu,Tong:2015xaa,Doroud:2015fsz},
many aspects of this subject still remain unexplored. 

One interesting general question is which backgrounds of Newton-Cartan
gravity admit non-relativistic supersymmetry, and how to systematically
construct Lagrangians on these backgrounds. In the relativistic case,
a systematic approach to this question was outlined by Festuccia and
Seiberg \cite{Festuccia:2011ws}. Starting with an off-shell formulation
of supergravity coupled to matter fields, one proceeds to take the
``rigid limit'' by freezing out graviton and gravitino fluctuations,
thereby obtaining a supersymmetric theory on a curved background. The conditions
for a background to be supersymmetric are found by demanding that the
gravitino variation vanishes. This in turn leads to Killing spinor
equations in curved space, which can be studied systematically to
classify supersymmetric backgrounds \cite{Jia:2011hw,Samtleben:2012gy,Klare:2012gn,Dumitrescu:2012ha,Liu:2012bi,Dumitrescu:2012at}.

In this paper, we initiate a similar approach to classifying curved
Newton-Cartan backgrounds that admit field theories with non-relativistic
supersymmetry. Starting with the off-shell version of three-dimensional
Newton-Cartan supergravity found in \cite{Bergshoeff:2015uaa}, we
proceed to decouple gravity. Demanding that the gravitino and its
variation vanish leads to a non-relativistic Killing spinor equation,
which we analyze in detail. Using integrability conditions, we can
derive the necessary and sufficient conditions for backgrounds to
admit four supercharges (unbroken supersymmetry), and also study examples of backgrounds with
reduced supersymmetry ($\frac{1}{2}$-BPS solutions). The supersymmetric
solutions found this way can be characterized by a ``gravitational
force'' field $\Phi_{i}(t,\vec{x})=\Gamma_{\phantom{i}00}^{i}$, and a
``Coriolis force'' field $C(t,\vec{x})=\frac{1}{2}\epsilon_{ij}\Gamma_{\phantom{}0j}^{i}$,
both of which represent the curvature induced by foliating the temporal
slices in a non-trivial way along the absolute time direction $\tau_{\mu}$.
Interestingly, a necessary condition for a background to preserve
any number of supersymmetry is that the spatial curvature, captured
by $\Gamma_{\phantom{i}jk}^{i}$, vanishes. 

Since backgrounds of Newton-Cartan gravity are formulated in a somewhat
unfamiliar language, using either two degenerate metrics, or one spatial
metric and a ``velocity'' field $\tau_{\mu}$, it is instructive
to connect our results to those for the relativistic ${\cal N}=2$
supergravity theory. Given that the Newton-Cartan supergravity theory of
\cite{Bergshoeff:2015uaa} was obtained as the $c\rightarrow\infty$
limit of the relativistic theory \cite{Howe:1995zm,Achucarro:1989gm}, one may ask whether the same limit
can be taken already at the level of the Killing spinor equations
themselves, in order to relate relativistic to non-relativistic 
backgrounds. Although taking this limit is possible, we are not guaranteed
to end up with the same BPS conditions for non-relativistic backgrounds
that we do by starting with Newton-Cartan supergravity, and freezing
out gravity (see figure \ref{fig:diagram2}). In other words, the rigid limit and the
$c\rightarrow\infty$ limit do not commute. The non-commutativity
is due to the additional constraints on auxiliary fields that are
imposed in Newton-Cartan supergravity, where the gravitino is generally
nonzero. These conditions are not needed at the level of rigid supersymmetry without gravity. 

The rest of this paper is organized as follows. In section \ref{sec:NC_gravity},
we briefly review Newton-Cartan gravity in both the metric and vielbein
formalism. In section \ref{sec:NC_sugra}, we review the off-shell
version of three-dimensional Newton-Cartan supergravity found in \cite{Bergshoeff:2015uaa},
and in particular the limiting procedure that was used to derive the theory.
In section \ref{sec:nonrel_back}, we derive and analyze the non-relativistic
Killing spinor equation. Using integrability conditions, we determine
all backgrounds with maximal supersymmetry, and also study examples
of backgrounds with reduced supersymmetry.
In section \ref{sec:rel_back}, we turn to the relativistic ${\cal{N}}=2$ supergravity
theory and study supersymmetric backgrounds using the same method
of rigid supersymmetry as before. Using integrability, we derive all solutions admitting four relativistic
supercharges. In section \ref{sec:limit}, we discuss the $c\rightarrow\infty$
limit of the relativistic Killing spinor equation, and show that it
leads to a bigger class of non-relativistic solutions than those found
in section \ref{sec:nonrel_back}. We conclude with a discussion of
our results and point towards some interesting future directions.

\begin{center}
\begin{figure}
\begin{centering}
\includegraphics[width=12cm]{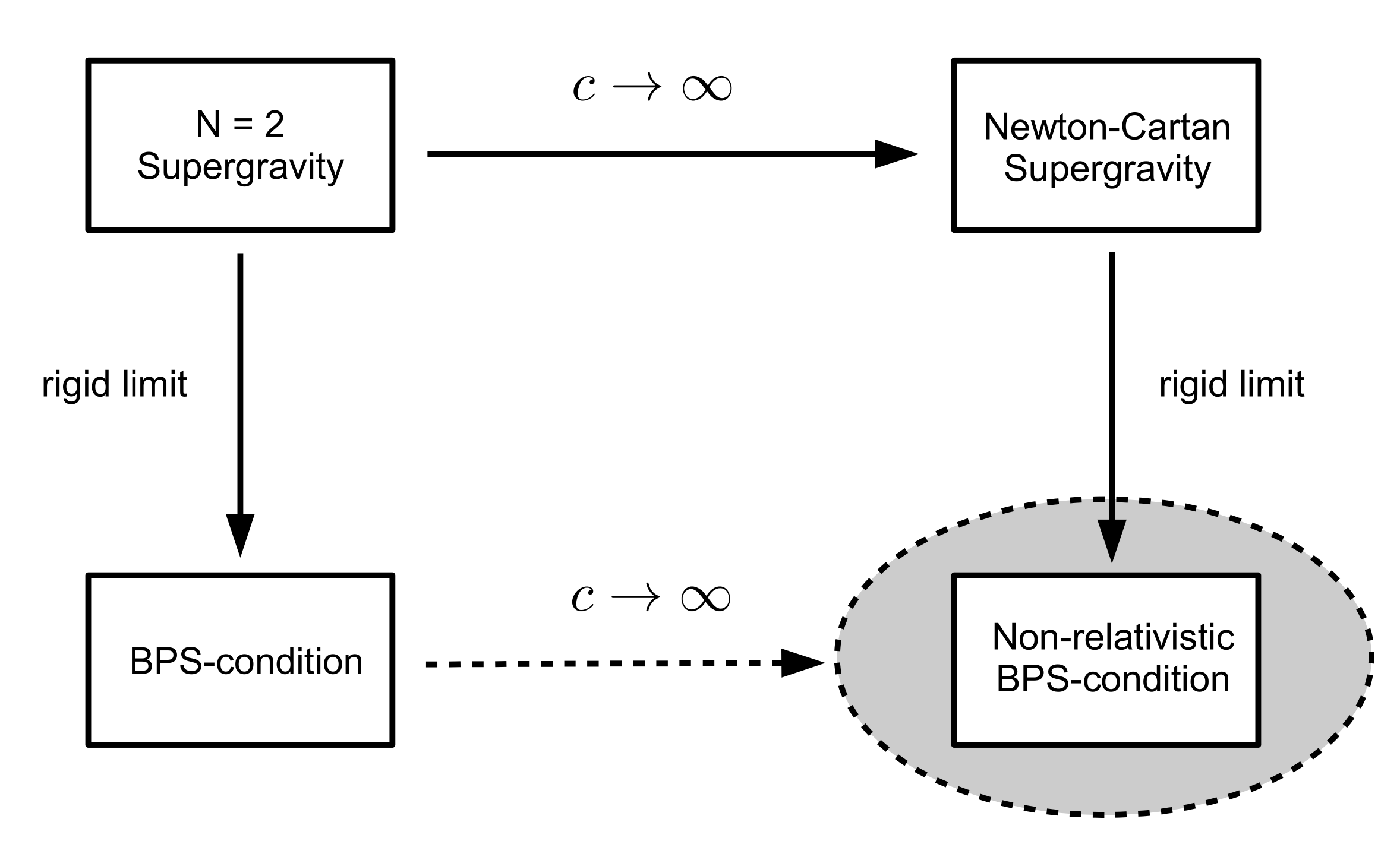}
\caption{\label{fig:diagram2}The difference between the non-relativistic BPS-condition, (\ref{eq:BPS_A}) through (\ref{eq:BPS_D}), obtained by taking the rigid limit of Newton-Cartan supergravity, and the set of backgrounds obtained by taking the non-relativistic limit of the relativistic supersymmetric solutions (grey).
In general, the latter is a superset of the former.}
\par\end{centering}
\end{figure}
\par\end{center}

\section{Newton-Cartan Gravity}

\label{sec:NC_gravity}To set the stage for our supergravity analysis,
let us first review Newton-Cartan gravity \cite{Misner:1974qy,Datcourt}. Newton-Cartan
gravity is a covariant formulation of Newtonian gravity. Due to its non-relativistic nature, this theory is commonly formulated
in terms of a temporal metric $\tau_{\mu\nu}$ , and a separate spatial
metric $h^{\mu\nu}$, making spatial and temporal distances two separate,
well-defined quantities. Both metrics are degenerate, which can be
understood heuristically by considering the example of the non-relativistic
limit of the Minkowski metric \cite{Andringa:2010it}:
\begin{equation}
\eta_{\mu\nu}=\left(\begin{array}{cc}
-c^{2} & 0\\
0 & I_{D-1}
\end{array}\right),\qquad\eta^{\mu\nu}=\left(\begin{array}{cc}
-1/c^{2} & 0\\
0 & I_{D-1}
\end{array}\right),
\end{equation}
In the limit $c\rightarrow\infty$, the metric naturally splits into
a temporal and a spatial metric: 
\begin{equation}
\eta_{\mu\nu}\rightarrow\tau_{\mu\nu},\qquad\eta^{\mu\nu}\rightarrow h^{\mu\nu},
\end{equation}
where in our case, $\tau_{\mu\nu}=-c^{2}\delta_{\mu}^{0}\delta_{\nu}^{0}$,
$h^{\mu\nu}=\delta^{ij}$. The degeneracy can be expressed as 
\begin{equation}
\tau_{\mu\rho}h^{\rho\nu}=0.\label{eq:degen}
\end{equation}
The temporal metric may be written as $\tau_{\mu\nu}=\tau_{\mu}\tau_{\nu}$.
Intuitively, we can understand the geometries within Newton-Cartan
gravity in the following way: The 1-form $\tau_{\mu}$ defines a
global time direction. At each moment in time, there is a Riemannian
space with (inverse) metric $h^{\mu\nu}$. The connection to Newtonian
gravity is established by choosing the curvature of the resulting manifold in
such a way that the geodesics of particles moving in the curved space
geometry are equivalent to the curved paths of classical particles
in flat space.

\subsection{Vielbein Formalism}
\label{sub:vielbein}
In this paper, we will also use an alternative formulation of Newton-Cartan
gravity in terms of vielbein fields \cite{Andringa:2010it}. Recall that General Relativity in $D$ dimensions
can be formulated as a gauge theory of the Poincar\'e algebra, which
has generators $P_{A}$ and $M_{AB}$ ($A,B=0,1,...,D-1)$. The associated
gauge fields are the vielbein $E_{\mu}^{A}$, and the spin connection
$\Omega_{\mu}^{AB}$. Newton-Cartan gravity can be written in the
same language. To accomplish this, we first identify the generators
of the non-relativistic symmetry group. In our case, the generators
are given by time translations $H$, spatial translations $P_{a}$,
rotations $J_{ab}$, and Galilean boosts $G_{a}$ ($a=1,...,D-1$), which together
form the Galilean algebra. To connect the relativistic and non-relativistic
symmetry groups, it turns out to be more natural to consider the Bargmann
algebra, which is the central extension of the Galilean algebra by
a $U(1)$ generator $Z$ \cite{Andringa:2010it}. The Bargmann algebra can be obtained
by performing an In\"on\"u-Wigner contraction of the Poincar\'e algebra
\cite{Bergshoeff:2015uaa}. For each of the generators, we then introduce corresponding
gauge fields: 
\begin{align}
&&\mbox{time translations:}\kern-4em&\quad&H&\leftrightarrow\tau_\mu&&&\label{eq:Gal1}\\
&&\mbox{spatial translations:}\kern-4em&\quad&P_a&\leftrightarrow e_\mu^a\\
&&\mbox{rotations:}\kern-4em&\quad&J_{ab}&\leftrightarrow\omega_\mu^{ab}\\
&&\mbox{Galilean boosts:}\kern-4em&\quad&G_a&\leftrightarrow\omega_\mu^a\label{eq:Gal4}\\
&&\mbox{U(1):}\kern-4em&\quad&Z&\leftrightarrow m_\mu
\end{align}
We see that the spacetime-translation generator $P_{A}$ of the Poincar\'e
algebra has split up into time translations $H$ and spatial translations
$P_{a}$. Correspondingly, the vielbein splits as $E_{\mu}^{A}\rightarrow(\tau_{\mu},e_{\mu}^{a})$, where $\tau_{\mu}$ is a ``temporal vielbein''
and $e_{\mu}^{a}$ is a ``spatial vielbein''. In a similar fashion,
the spin connection $\Omega_{\mu}^{AB}$ splits up into an $SO(2)$
spin-connection $\omega_{\mu}^{ab}$, and a boost connection $\omega_{\mu}^{a}$.
Finally, the abelian gauge field $Z$ provides a central extension
of the Galilean algebra (\ref{eq:Gal1})-(\ref{eq:Gal4}) to the full
Bargmann algebra. It is needed to consistently perform the contraction
of the Poincar\'e algebra. Later we will see that the geometric
role of the corresponding gauge field $m_{\mu}$ is to define the
rules of parallel transport, or equivalently to define a connection
$\Gamma$. 

Next, we define inverse vielbein fields $e_{a}^{\mu}$, $\tau^{\mu}$
such that
\begin{align}
e_{a}^{\mu}e_{\mu}^{b} & =\delta_{a}^{b}, &  &  & \tau^{\mu}\tau_{\mu} & =1,\label{eq:e_tau_constr1}\\
e_{a}^{\rho}e_{\mu}^{a} & =\delta_{\mu}^{\rho}-\tau_{\mu}\tau^{\rho}, &  &  & \tau^{\mu}e_{\mu}^{a} & =\tau_{\mu}e_{a}^{\mu}=0.\label{eq:e_tau_constr_2}
\end{align}
Note that flat spatial indices $a,b,...$ are contracted using $\delta^{ab}$.
The degenerate metrics introduced previously are given in terms of
the vielbeine in the usual way:
\begin{equation}
h_{\mu\nu}=e_{\mu}^{a}e_{\nu a},\qquad h^{\mu\nu}=e_{a}^{\mu}e^{\nu a},\qquad\tau_{\mu\nu}=\tau_{\mu}\tau_{\nu}.
\end{equation}
The constraints (\ref{eq:e_tau_constr_2}) then imply

\begin{equation}
h^{\mu\nu}h_{\nu\rho}=\delta_{\mu}^{\rho}-\tau_{\mu}\tau^{\rho},\qquad h_{\mu\nu}\tau^{\nu}=h^{\mu\nu}\tau_{\nu}=0.\label{eq:h_tau_constr}
\end{equation}
We will make use of these conditions when constructing explicit background
metrics later. Although the vielbein formalism is useful to construct
Newton-Cartan (super)gravity, we will also use the metric $h^{\mu\nu}$
when studying particular backgrounds, as it connects more directly
to the familiar metric formulation of General Relativity.

\subsection{Constraints and Adapted Coordinates}
\label{sub:adapted}
We can define gauge covariant curvatures of each of the gauge fields.
In the relativistic (Poincar\'e) case, those curvatures are given by
\begin{align}
R_{\mu\nu}^{\phantom{\mu\nu}A}(E) & =2\partial_{[\mu}E_{\nu]}^{A}-2\Omega_{[\mu}^{AB}E_{\nu]B},\\
R_{\mu\nu}^{\phantom{\mu\nu}AB}(\Omega) & =2\partial_{[\mu}\Omega_{\nu]}^{AB}-2\Omega_{[\mu}^{AC}\Omega_{\nu]C}^{\phantom{\nu]C}B}.
\end{align}
By imposing the first structure equation 
\begin{equation}
R_{\mu\nu}^{\phantom{\mu\nu}A}(E)=0,\label{eq:torsionfree}
\end{equation}
we can solve for the spin connection $\Omega_{\mu}^{AB}$ in terms
of $E_{\mu}^{A}.$ In complete analogy with the relativistic case,
we can define gauge covariant curvatures corresponding to each of
the generators in Newton-Cartan gravity:
\begin{alignat}{1}
R_{\mu\nu}(H) & =2\partial_{[\mu}\tau_{\nu]},\label{eq:NC_curv_H}\\
R_{\mu\nu}^{\phantom{\mu\nu}a}(P) & =2\partial_{[\mu}e_{\nu]}^{a}-2\omega_{[\mu}^{ab}e_{\nu]b}^{\phantom{\nu]}}-2\omega_{[\mu}^{a}\tau_{\nu]},\\
R_{\mu\nu}^{\phantom{\mu\nu}ab}(J) & =2\partial_{[\mu}\omega_{\nu]}^{ab}-2\omega_{[\mu}^{ac}\omega_{\nu]c}^{\phantom{\nu]c}b},\\
R_{\mu\nu}^{\phantom{\mu\nu}a}(G) & =2\partial_{[\mu}\omega_{\nu]}^{a}-2\omega_{[\mu}^{ab}\omega_{\nu]b},\\
R_{\mu\nu}(Z) & =2\partial_{[\mu}m_{\nu]}-2\omega_{[\mu}^{a}e_{\nu]a}.\label{eq:NC_curv_Z}
\end{alignat}
In the absence of additional structure,
one can show that taking the non-relativistic limit of the torsion-free
condition (\ref{eq:torsionfree}) consistently requires imposing the following
constraints on the non-relativistic curvature tensors \cite{Bergshoeff:2015uaa}:
\begin{align}
R_{\mu\nu}(H) & =R_{\mu\nu}^{\phantom{\mu\nu}a}(P)=R_{\mu\nu}(Z)=0. \label{eq:NR_curv_identities}
\end{align}
These constraints can be used to determine the connections $\omega_{\mu}^{ab}$
and $\omega_{\mu}^{a}$ in terms of $(\tau_{\mu},e_{\mu}^{a},m_{\mu})$:
\begin{align}
\omega_{\mu}^{ab} & =-2e^{\nu[a}\partial_{[\mu}e_{\nu]}^{b]}+e_{\mu}^{c}e^{\rho a}e^{\nu b}\partial_{[\rho}e_{\nu]}^{c}-\tau_{\mu}e^{\rho a}e^{\nu b}\partial_{[\rho}m_{\nu]},\\
\omega_{\mu}^{a} & =\tau^{\nu}\partial_{[\mu}e_{\nu]}^{a}+e_{\mu}^{b}e^{\rho a}\tau^{\nu}\partial_{[\rho}e_{\nu]}^{b}+e^{\nu a}\partial_{[\mu}m_{\nu]}-\tau_{\mu}e^{\rho a}\tau^{\nu}\partial_{[\rho}m_{\nu]}.
\end{align}
A background of Newton-Cartan gravity is therefore uniquely determined
by choosing $(\tau_{\mu},e_{\mu}^{a},m_{\mu})$.

The constraint $R_{\mu\nu}(H)=2\partial_{[\mu}\tau_{\nu]}=0$ gives rise to torsionless Newton-Cartan
gravity.  In fact, this torsion-free condition can be relaxed by including a background gauge
field while taking the non-relativistic limit \cite{Jensen:2014wha}.  For our present analysis, however, we restrict to
the torsionless theory, in which case we can locally write $\tau_{\mu}=\partial_{\mu}T(x^{\nu})$. By construction,
$\tau_{\mu}$ singles out a time-direction, so it is useful to introduce
``adapted coordinates'' by letting $T(x^{\mu})=x^{0}\equiv t$,
so that $\tau_{\mu}=\delta_{\mu}^{0}$. In these coordinates, the
constraints (\ref{eq:e_tau_constr1}) and (\ref{eq:h_tau_constr})
imply 
\begin{align}
\tau_{\mu} & =\delta_{\mu}^{0},\\
\tau^{\mu} & =(1,v^{i}),\\
h^{\mu0} & =0,\\
h_{\mu0} & =-h_{\mu i}v^{i}.
\end{align}
Note that our gauge choice does not completely fix the coordinates.
In fact, there is a residual gauge freedom given by
\begin{align}
t & \rightarrow t+\mathrm{const.},\nonumber \\
x^{i} & \rightarrow F^{i}(t,x^{j}),\label{eq:gauge}
\end{align}
with $\mathrm{det}\frac{\partial F^{i}}{\partial x^{j}}\neq0$.

\subsection{Connection and Interpretation of $m_{\mu}$}

Given a background characterized by $(\tau_{\mu},e_{\mu}^{a},m_{\mu})$,
we can uniquely define a connection by imposing the vielbein postulate
\cite{Andringa:2010it}
\begin{align}
\partial_{\mu}e_{\nu}^{a}-\omega_{\mu}^{ab}e_{\nu b}-\omega_{\mu}^{a}\tau_{\nu}-\Gamma_{\phantom{\rho}\nu\mu}^{\rho}e_{\rho}^{a} & =0,\\
\partial_{\mu}\tau_{\nu}-\Gamma_{\phantom{\lambda}\nu\mu}^{\lambda}\tau_{\lambda} & =0,
\end{align}
which can be solved to find the connection coefficients
\begin{align}
\Gamma_{\phantom{\rho}\mu\nu}^{\rho} & =\tau^{\rho}\partial_{(\mu}\tau_{\nu)}+e_{a}^{\rho}(\partial_{(\mu}e_{\nu)}^{a}-\omega_{(\mu}^{ab}e_{\nu)b}-\omega_{(\mu}^{a}\tau_{\nu)})\nonumber \\
 & =\tau^{\rho}\partial_{(\mu}\tau_{\nu)}+\ft12h^{\rho\lambda}(\partial_{\nu}h_{\lambda\mu}+\partial_{\mu}h_{\lambda\nu}-\partial_{\lambda}h_{\mu\nu}+2K_{\lambda(\mu}\tau_{\nu)}),
\end{align}
where 
\begin{equation}
K_{\mu\nu}=2\partial_{[\mu}m_{\nu]}.
\end{equation}
As we will see, the definition of $\Gamma$ is the only place where
$m_{\mu}$ shows up. Therefore, $m_{\mu}$ plays the role of determining
the rules of parallel transport in a given background $(\tau_{\mu},e_{\mu}^{a})$. 

The Riemann tensor can be written in terms of the connection in the
usual way:
\begin{align}
R_{\phantom{\mu}\nu\rho\sigma}^{\mu}(\Gamma) & =\partial_{\rho}\Gamma_{\phantom{\mu}\nu\sigma}^{\mu}-\partial_{\sigma}\Gamma_{\phantom{\mu}\nu\rho}^{\mu}+\Gamma_{\phantom{\lambda}\nu\sigma}^{\lambda}\Gamma_{\phantom{\mu}\lambda\rho}^{\mu}-\Gamma_{\phantom{\lambda}\nu\rho}^{\lambda}\Gamma_{\phantom{\mu}\lambda\sigma}^{\mu}\label{eq:riemann}
\end{align}
Alternatively, we can express it in terms of the boost- and spin-curvature
tensors of the Bargmann algebra:
\begin{equation}
R_{\phantom{\mu}\nu\rho\sigma}^{\mu}(\Gamma)=-e_{a}^{\mu}\left(R_{\rho\sigma}^{\phantom{\rho\sigma}a}(G)\tau_{\nu}+R_{\rho\sigma}^{\phantom{\rho\sigma}ab}(J)e_{\nu b}\right).
\end{equation}

\section{Off-shell Newton-Cartan Supergravity}

\label{sec:NC_sugra}

Three-dimensional Newton-Cartan supergravity
can be constructed as a gauge theory of the supersymmetric extension
of the Bargmann algebra introduced in section \ref{sub:vielbein} \cite{Andringa:2013mma}. To derive an
off-shell version of this theory, one starts with an off-shell realization
of $\mathcal{N}=2$ supergravity and performs an In\"on\"u-Wigner contraction
that reduces the relativistic supersymmetry algebra to the super-Bargmann
algebra \cite{Bergshoeff:2015uaa}. 

In three dimensions, there are two inequivalent formulations of ${\cal N}=2$ supergravity,
namely the ${\cal N}=(1,1)$ theory \cite{Rocek:1985bk,Nishino:1991sr,Howe:1995zm} and the ${\cal N}=(2,0)$ theory \cite{Howe:1995zm}. 
We will focus on the $(2,0)$ theory, since it was used as a starting point for 
constructing the torsionless Newton-Cartan supergravity 
theory of \cite{Bergshoeff:2015uaa}.
The gravity multiplet of both ${\cal{N}}=2$ supergravity theories
contains a vielbein $E_{\mu}^{A}$ $(A=0,1,2)$ and two gravitini
$\Psi_{\mu i}$ $(i=1,2)$, which are Majorana spinors with two real
components each. The off-shell multiplet of the $(2,0)$ theory additionally contains
two gauge fields $M_{\mu}$ and $V_{\mu}$, as well as a scalar
$D$. The variations of each field under supersymmetry can be found
in \cite{Howe:1995zm,Bergshoeff:2015uaa}. Here we focus only on the transformation properties of
the gravitino. Under a combined supersymmetry transformation (parametrized
by two Majorana spinors $\eta_{i}$) and $U(1)_R$-transformation (parametrized
by $\rho$), the gravitino transforms as 
\begin{equation} \label{eq:gravitinovariation}
\delta\Psi_{\mu i}=\nabla_{\mu}\eta_{i}+\epsilon^{ij}\eta_{j}V_{\mu}-\gamma_{\mu}\eta_{i}D+\ft14\gamma_{\mu}\gamma_{\rho\sigma}\hat{F}^{\rho\sigma}\epsilon^{ij}\eta_{j}-\epsilon^{ij}\Psi_{\mu j}\rho,
\end{equation}
where 
\begin{equation}
\hat{F}_{\mu\nu}=2\partial_{[\mu}M_{\nu]}-\ft12\epsilon^{ij}\bar{\Psi}_{[\mu i}\Psi_{\nu]j},
\end{equation}
and $\nabla_{\mu}=\partial_{\mu}-\ft14\Omega_{\mu}^{AB}\gamma_{AB}$. 

Off-shell Newton-Cartan supergravity is constructed by taking a non-relativistic
limit of the fields that mirrors the limit taken in the contraction
of the Poincar\'e algebra. Let us give a brief review of this limiting
procedure, as outlined in \cite{Bergshoeff:2015uaa}. One starts by redefining the bosonic
fields as follows:
\begin{align} \label{eq:omegaexpansion}
E_{\mu}^{A} & =\delta_{0}^{A}(\omega\tau_{\mu}+\frac{1}{2\omega}m_{\mu})+\delta_{a}^{A}e_{\mu}^{a},\nonumber \\
M_{\mu} & =\omega\tau_{\mu}-\frac{1}{2\omega}m_{\mu},\nonumber \\
D & =\frac{1}{\omega}S,
\end{align}
where $a=1,2$. The spinors are first rewritten as
\begin{equation} \label{eq:plusminusbasis}
\Psi_{\pm}=\frac{1}{\sqrt{2}}\left(\Psi_{1}\pm\gamma_{0}\Psi_{2}\right),\qquad\eta_{\pm}=\frac{1}{\sqrt{2}}\left(\eta_{1}\pm\gamma_{0}\eta_{2}\right),
\end{equation}
and then rescaled according to 
\begin{align} \label{eq:spinorrescaling} 
\Psi_{+} & =\sqrt{\omega}\psi_{+},\qquad & \eta_{+} & =\sqrt{\omega}\epsilon_{+},\nonumber \\
\Psi_{-} & =\frac{1}{\sqrt{\omega}}\psi_{-},\qquad & \eta_{-} & =\frac{1}{\sqrt{\omega}}\epsilon_{-}.
\end{align}
Finally, the curvature form splits into spatial and temporal components
as follows:
\begin{eqnarray}
\Omega_{\mu}^{ab} & = & \omega_{\mu}^{ab}+O\bigg(\frac{1}{\omega^{2}}\bigg),\label{eq:spin_conn}\\
\Omega_{\mu}^{0a} & = & \frac{1}{\omega}\omega_{\mu}^{a}+O\bigg(\frac{1}{\omega^{3}}\bigg).\label{eq:boost_conn}
\end{eqnarray}
The one-form $\omega_{\mu}^{a}$ is a boost connection, while $\omega_{\mu}^{ab}$
is a spin connection for spatial rotations.

Next, we take the limit $\omega\rightarrow\infty$, which can be thought
of as taking $c\rightarrow\infty.$ To eliminate divergences that
appear in the transformation laws, one is forced to impose the following
constraints on the bosonic fields:  
\begin{align}
\partial_{[\mu}\tau_{\nu]} & =0,\label{eq:constraint_tau}\\
V_{\mu} & =-2\tau_{\mu}S,\label{eq:constraint_V}\\
\hat{F}_{\mu\nu} & =2\partial_{[\mu}M_{\nu]}-\ft12\epsilon^{ij}\bar{\Psi}_{[\mu i}\Psi_{\nu]j}=0. \label{eq:constraint_F}
\end{align}
Finally, the non-relativistic variations of the gravitini take the
form
\begin{align}
\delta\psi_{\mu+} & =D_{\mu}\epsilon_{+}+S\tau_{\mu}\gamma_{0}\epsilon_{+}+\gamma_{0}\psi_{\mu+}\rho,\label{eq:del_psi_plus}\\
\delta\psi_{\mu-} & =D_{\mu}\epsilon_{-}-3S\tau_{\mu}\gamma_{0}\epsilon_{-}+\ft12\omega_{\mu}^{a}\gamma_{a0}\epsilon_{+}-Se_{\mu}^{a}\gamma_{a}\epsilon_{+}-\gamma_{0}\psi_{\mu-}\rho,\label{eq:del_psi_minus}
\end{align}
where the derivative operator $D_{\mu}\equiv\partial_{\mu}-\frac{1}{4}\omega_{\mu}^{ab}\gamma_{ab}$
is covariant under local spatial rotations. The non-relativistic supergravity
multiplet consists of $\tau_{\mu}$,$e_{\mu}^{a}$, $\psi_{\mu\pm}$,
as well as the auxiliary fields $m_{\mu}$ and $S$. The vielbein
$E_{\mu}^{A}$ has split up into a temporal vielbein $\tau_{\mu}$
and a separate spatial vielbein $e_{\mu}^{a}$ (see section \ref{sub:vielbein}).
Finally, note that this construction gives rise to torsionless Newton-Cartan supergravity.  The
constraints (\ref{eq:constraint_tau}), (\ref{eq:constraint_V}) and (\ref{eq:constraint_F}) can be lifted
by instead considering the torsionful theory \cite{Bergshoeff:2015ija}.

\section{Non-relativistic Supersymmetric Backgrounds}

\label{sec:nonrel_back}To find backgrounds that respect non-relativistic
supersymmetry, we proceed by demanding that the gravitini $\psi_{\pm}$
and their variations (\ref{eq:del_psi_plus}) and (\ref{eq:del_psi_minus})
vanish. This guarantees that the bosonic fields do not vary under
supersymmetry, and in addition gives rise to the following Killing
spinor equations:
\begin{eqnarray}
D_{\mu}\epsilon_{+} & = & -S\tau_{\mu}\gamma_{0}\epsilon_{+},\label{eq:KSE1}\\
D_{\mu}\epsilon_{-} & = & 3S\tau_{\mu}\gamma_{0}\epsilon_{-}-\ft12\omega_{\mu}^{a}\gamma_{a0}\epsilon_{+}+Se_{\mu}^{a}\gamma_{a}\epsilon_{+}.\label{eq:KSE2}
\end{eqnarray} 
Each solution $(\epsilon_{+},\epsilon_{-})$ of the equations above
corresponds to a single preserved supercharge. To determine when
such a solution exists, we examine the integrability conditions 
\begin{equation}
\left[D_{\mu,}D_{\nu}\right]\epsilon_{\pm}=0.
\end{equation}
Using $\left[D_{\mu,}D_{\nu}\right]\epsilon_{\pm}=-\frac{1}{4}R_{\mu\nu}^{\phantom{\mu\nu}ab}(J)\gamma_{ab}\epsilon_{\pm}$,
(\ref{eq:KSE1}) and (\ref{eq:KSE2}), the integrability
conditions take the form 
\begin{align}
A_{\mu\nu}\gamma_{0}\epsilon_{+} & =0,\nonumber \\
B_{\mu\nu}\gamma_{0}\epsilon_{-}+C_{\mu\nu}^{a}\gamma_{a}\epsilon_{+} & =0,\label{eq:nr_integrability}
\end{align}
where
\begin{align}
A_{\mu\nu} & =-\ft14R_{\mu\nu}^{\phantom{\mu\nu}ab}(J)\epsilon_{ab}-2\tau_{[\mu}\partial_{\nu]}S,\label{eq:int_A}\\
B_{\mu\nu} & =-\ft14R_{\mu\nu}^{\phantom{\mu\nu}ab}(J)\epsilon_{ab}+6\tau_{[\mu}\partial_{\nu]}S,\label{eq:int_B}\\
C_{\mu\nu}^{a} & =-\ft12\epsilon_{\phantom{a}b}^{a}R_{\mu\nu}^{\phantom{\mu\nu}b}(G)+2e_{[\mu}^{a}\partial_{\nu]}S-4S^{2}\epsilon_{\phantom{a}b}^{a}e_{[\mu}^{b}\tau_{\nu]}.\label{eq:int_C}
\end{align}
To arrive at these expressions, we have used the constraints (\ref{eq:NR_curv_identities}). 

Assuming that $(\epsilon_{+},\epsilon_{-})$ span a 4-dimensional
spinor space, the necessary and sufficient condition for integrability
is $A_{\mu\nu}=B_{\mu\nu}=C_{\mu\nu}^{a}=0$. This is the maximally
supersymmetric case with four supercharges, which we analyze
further in section \ref{sub:4sc}. 

To find backgrounds with less than maximal supersymmetry, one may
consider imposing further constraints on the Killing spinors, e.g.
$\epsilon_{-}=0$. In this case, however, the integrability condition
needs to be rederived in the appropriate lower-dimensional subspace
of solutions, and may take a different form%
\footnote{For example, it is easy to convince oneself that setting $\epsilon_{-}=0$,
$A_{\mu\nu}=C_{\mu\nu}^{a}=0$ solves (\ref{eq:nr_integrability}), but
plugging this ansatz back into (\ref{eq:KSE1}) and (\ref{eq:KSE2})
does not guarantee a solution.%
}. We will study examples of such $\frac{1}{2}$-BPS backgrounds
with two supercharges in section \ref{sub:1/2BPS}. 

As we will demonstrate later, the existence of a single supercharge
implies the existence of at least one more supercharge, i.e. solutions
to the Killing spinor equations always come in pairs. Hence there
are no $\frac{1}{4}$-BPS solutions.

\subsection{Maximally Supersymmetric Solutions}
\label{sub:4sc}Backgrounds with completely unbroken supersymmetry admit four real supercharges,
or equivalently four linearly independent Killing spinors of the form
$(\epsilon_{+},\epsilon_{-})$. To solve the integrability condition
(\ref{eq:nr_integrability}), we therefore need to demand $A_{\mu\nu}=B_{\mu\nu}=C_{\mu\nu}^{a}=0$,
which implies
\begin{align}
\tau_{[\mu}\partial_{\nu]}S & =0,\label{eq:int_max_1}\\
R_{\mu\nu}^{\phantom{\mu\nu}ab}(J) & =0,\label{eq:int_max_2}\\
R_{\mu\nu}^{\phantom{\mu\nu}a}(G) & =8S^{2}\tau_{[\mu}e_{\nu]}^{a}-4\epsilon_{\phantom{a}b}^{a}e_{[\mu}^{b}\partial_{\nu]}S.\label{eq:int_max_3}
\end{align}
Together with the constraint $R_{\mu\nu}(H)=2\partial_{[\mu}\tau_{\nu]}=0$ (see section \ref{sub:adapted}),
these equations completely determine the maximally supersymmetric
backgrounds. To make contact with the more familiar language of General
Relativity, it is useful to translate the constraints (\ref{eq:int_max_2})
and (\ref{eq:int_max_3}) into conditions on the Riemann tensor constructed
from the Christoffel connection, 
\begin{equation}
R_{\phantom{\mu}\nu\rho\sigma}^{\mu}(\Gamma)=-e_{a}^{\mu}\left(\tau_{\nu}R_{\rho\sigma}^{\phantom{\rho\sigma}a}(G)+e_{\nu b}R_{\rho\sigma}^{\phantom{\rho\sigma}ab}(J)\right).\label{eq:Riem_Gamma}
\end{equation}
Using this expression, we can rewrite the conditions (\ref{eq:int_max_1})
through (\ref{eq:int_max_3}) as 
\begin{align}
\tau_{[\mu}\partial_{\nu]}S & =0,\label{eq:int_max_re_1}\\
R_{\phantom{\mu}\nu\rho\sigma}^{\mu}(\Gamma) & =-8S^{2}\tau_{\nu}\tau_{[\rho}\delta_{\sigma]}^{\mu}+4\epsilon_{\phantom{a}b}^{a}e_{a}^{\mu}\tau_{\nu}e_{[\rho}^{b}\partial_{\sigma]}S.\label{eq:int_max_re_2}
\end{align}
The Ricci tensor is given by 
\begin{equation}
R_{\mu\nu}=8S^{2}\tau_{\mu}\tau_{\nu}.\label{eq:Ricci_constraint} 
\end{equation}
This is, in fact, the standard Einstein equation for Newton-Cartan gravity,
$R_{\mu\nu}=4\pi G\rho\tau_\mu\tau_\nu$
with $S^2$ playing the role of the Newtonian mass density $\rho$.  Note, however, that here it arises as a
condition of maximal supersymmetry, and not through the direct imposition of any equations of motion.

To analyze (\ref{eq:int_max_re_1}) and (\ref{eq:int_max_re_2}) further,
we introduce adapted coordinates (see section \ref{sub:adapted}).
In these coordinates, the first of the two constraints simply becomes
\begin{equation}
\partial_{i}S=0.
\end{equation}
To evaluate the second constraint, we first note that (\ref{eq:Ricci_constraint})
implies  $R_{ij}=0$. The spatial metric $h_{ij}$ is therefore flat, with a
possible time dependence: 
\begin{equation}
h_{ij}=g(t)\delta_{ij}.
\end{equation}
We can use the gauge freedom (\ref{eq:gauge}) to set 
\begin{align}
h_{ij} & =\delta_{ij},\qquad h^{ij}=\delta^{ij}.
\end{align}
After making this gauge choice, the remaining allowed coordinate transformations
are
\begin{align}
t & \rightarrow t+\mathrm{const.},\nonumber \\
x^{i} & \rightarrow A_{\phantom{i}j}^{i}(t)x^{j}+a^{i}(t),\label{eq:Galilei}
\end{align}
where $A_{\phantom{i}k}^{i}A_{j}^{\phantom{j}k}=\delta_{j}^{i}$.
The $A_{\phantom{i}j}^{i}$ parametrize time-dependent rotations,
while $a^{i}(t)$ corresponds to a Galilean boost. For this reason,
the gauge choice $h_{ij}=\delta_{ij}$ is sometimes referred to as
choosing ``Galilean coordinates'' \cite{Datcourt}. In these coordinates, the conditions
(\ref{eq:h_tau_constr}) determine the spatial metric to take the form 
\begin{equation}
h_{\mu\nu}=\left(\begin{array}{cc}
v^{i}v_{i} & -v^{i}\\
-v^{i} & \delta_{ij}
\end{array}\right).\label{eq:h_explicit}
\end{equation}
Knowing the form of the spatial metric, we can explicitly write down the vielbein
and its inverse:
\begin{align}
e_{\mu}^{a} & =(-v^{a},\delta_{i}^{a}),\label{eq:vielb_flat1}\\
e_{a}^{\mu} & =(0,\delta_{a}^{i})^{T}.\label{eq:vielb_flat2}
\end{align}
We are now ready to write the second integrability condition (\ref{eq:int_max_re_2})
in Galilean coordinates. To express the Riemann tensor explicitly
in terms of metric components, first note that the only nonzero connection
coefficients are
\begin{align}
\Gamma_{\phantom{i}00}^{i} & =\partial_{i}(m_{0}-\ft12h_{00})-\partial_{0}(m_{i}-h_{i0})\equiv\Phi_{i},\label{eq:Phi_def}\\
\Gamma_{\phantom{i}0j}^{i} & =-\Gamma_{\phantom{i}0i}^{j}=\partial_{[i}(m_{j]}-h_{j]0})\equiv C_{ij}=-C_{ji}.\label{eq:C_def}
\end{align}
To simplify our discussion, it will be useful to define $C\equiv\frac{1}{2}\epsilon_{ij}C^{ij}$, with $\epsilon _{12}=1$.
Using these definitions, the Riemann-constraint (\ref{eq:int_max_re_2})
may be written as 
\begin{align}
\partial_{i}C & =0,\\
\partial_{j}\Phi_{i}-\epsilon_{ij}\partial_{0}C+\delta_{ij}C^{2} & =4\delta_{ij}S^{2}+2\epsilon_{ij}\partial_{0}S.
\end{align}
Taking the antisymmetric part of the second equation and using the
definitions (\ref{eq:Phi_def}) and (\ref{eq:C_def}), we find $\partial_{0}S=0$,
and therefore conclude that $S=\mathrm{const.}$

To summarize, backgrounds
admitting four supercharges are given by a degenerate spatial metric
$h_{\mu\nu}$ of the form (\ref{eq:h_explicit}), and connection coefficients
$\Phi_{i}$, $C$, such that 
\begin{align}
\partial_{(i}\Phi_{j)}+\delta_{ij}\left(C^{2}-4S^2\right) & =0,\label{eq:max_cond_2}\\
\partial_{i}C & =0,\label{eq:max_cond_1}\\
S & =\mathrm{const.} \label{eq:max_cond_3}
\end{align}
Given a specific background, the auxiliary field $S$ required to
close the non-relativistic SUSY algebra is found by solving (\ref{eq:max_cond_2}).
Since $S$ is constant, (\ref{eq:Ricci_constraint}) demonstrates that maximally supersymmetric
solutions are essentially Newtonian cosmologies with a homogeneous matter distribution
$\rho=2S^2/\pi G\ge0$.

\subsubsection{Connection Coefficients}

To give a physical interpretation to the connection coefficients $\Phi_{i}$
and $C$, let us consider the geodesic equation in the backgrounds
discussed above:
\begin{equation}
\frac{d^{2}x^{i}}{dt^{2}}+\Phi_{i}+2C_{ij}\frac{d{x}^{j}}{dt}=0,\label{eq:geodesic}
\end{equation}
We see that $\Phi_{i}$ represents the gravitational force, while
$C_{ij}$ is akin to the Coriolis force in a rotating reference frame.
Defining a scalar and vector potential via
\begin{align} \label{eq:phiAtomh}
\varphi & =m_{0}-\ft12h_{00},\\ \nonumber
A_{i} & =m_{i}-h_{i0},
\end{align}
we may use (\ref{eq:Phi_def}) and (\ref{eq:C_def}) to write the two force fields as 
\begin{eqnarray}
\Phi_{i} & = & \partial_{i}\varphi-\partial_{0}A_{i},\nonumber \\
C_{ij} & = & \partial_{[i}A_{j]}.\label{eq:phiC_potentials}
\end{eqnarray}
One may then identify $\Phi_{i}$
and $C_{ij}$ in (\ref{eq:geodesic}) as ``electric'' and ``magnetic'' - type fields,
which are invariant under the gauge transformation $\varphi\rightarrow\varphi+\partial_{0}\lambda$,
$A_{i}\rightarrow A_{i}+\partial_{i}\lambda$. 

Finally, note that the vector field $m_{\mu}$ is not part of the
metric itself, but only shows up in the expressions (\ref{eq:Phi_def}) and (\ref{eq:C_def}) for the connection
coefficients in a given background $h_{\mu\nu}$. Changing $m_{\mu}$
is equivalent to changing the rules for parallel transport in a fixed
background. 

We can solve the conditions (\ref{eq:max_cond_2}) through (\ref{eq:max_cond_3})
explicitly by performing a Galilei transformation (\ref{eq:Galilei})
into a non-rotating coordinate frame, where $C=0$. In this case $A_{i}$
is rotation-free, and we can locally write $A_{i}=\partial_{i}\psi$.
The constraint (\ref{eq:max_cond_2}) then takes the form
\begin{equation}
\partial_{i}\partial_{j}\hat{\varphi}=4S^{2}\delta_{ij},\label{eq:pseudo_poisson}
\end{equation}
where we introduced a new potential $\hat{\varphi}=\varphi-\partial_{0}\psi$.
Taking the trace of this equation, we recover Poisson's equation
with a source $\rho={2S^{2}}/{\pi G}=\mathrm{const.}$ However,
since (\ref{eq:pseudo_poisson}) also contains the additional condition
$\left(\partial_{1}^{2}-\partial_{2}^{2}\right)\hat{\varphi}=\partial_{1}\partial_{2}\hat{\varphi}=0$,
the solution is further constrained. We find
\begin{equation}
\hat{\varphi}(x,y)=2S^{2}(x^{2}+y^{2})+c_{1}(t)x+c_{2}(t)y+d(t),
\end{equation}
instead of the usual logarithmic solution for a gravitational potential of a homogeneous matter distribution.
A particle moving along a geodesic in a maximally supersymmetric background experiences
a Newtonian gravitational force 
\begin{equation}
\Phi_{i}=\partial_{i}\hat{\varphi}=4S^{2}x^{i}+c_{i}(t).
\end{equation}

\subsubsection{Killing Spinors}

We can explicitly construct all four supercharges of the
maximally supersymmetric case by solving (\ref{eq:KSE1}) and (\ref{eq:KSE2})
in the backgrounds constructed above. For backgrounds with $h_{ij}=\delta_{ij}$,
the Killing spinor equations take the form
\begin{align}
0 & =\partial_{i}\epsilon_{+},\\
0 & =\big[\partial_{0}+(\ft12C+S)\gamma_{0}\big]\epsilon_{+},\\
0 & =\partial_{i}\epsilon_{-}-(\ft12C+S)\gamma_{i}\epsilon_{+}-\ft12\epsilon_{ab}\partial_{i}v_{a}\gamma_{b}\epsilon_{+},\\
0 & =\big[\partial_{0}+(\ft12C-3S)\gamma_{0}\big]\epsilon_{-}+\big(S-\ft12C\big)v^{a}\gamma_{a}\epsilon_{+}-\ft12\epsilon_{ab}(\Phi_{a}+\partial_{0}v_{b})\gamma_{b}\epsilon_{+}.
\end{align}
These equations can be solved, provided the integrability conditions
found previously hold. The condition $\partial_{\mu}S=\partial_{i}C=0$
guarantees the existence of two linearly independent homogeneous solutions
\begin{equation}
\epsilon_{+}=0,\qquad\epsilon_{-}=e^{-\int dt^\prime(3S-\ft12C)\gamma_{0}}\epsilon_{0},
\end{equation}
with $\epsilon_{0}$ an arbitrary constant Majorana spinor. Using
(\ref{eq:max_cond_2}), one can show that there are two additional
inhomogeneous solutions:
\begin{equation}
\epsilon_{+}=e^{-\int dt^{\prime}(\ft12C+S)\gamma_{0}}\epsilon_{0}^{\prime},\qquad\epsilon_{-}=e^{\int dt^{\prime}(3S-\ft12C)\gamma_{0}}M(t)\epsilon_{0}^{\prime}.
\end{equation}
Here $\epsilon_{0}^{\prime}$ is another constant Majorana spinor,
and we defined
\begin{equation}
M(t)=\int^{t}dt^{\prime}\left[e^{\int^{t^{\prime}}dt^{\prime\prime}(\ft12C-3S)\gamma_{0}}\left((\ft12C-S)v^{a}\gamma_{a}+\ft12\epsilon_{ab}(\Phi_{a}+\partial_0{v}_{a})\gamma_{b}\right)e^{-\int^{t^{\prime}}dt^{\prime\prime}(\ft12C+S)\gamma_{0}}\right].
\end{equation}
This concludes our discussion of maximally supersymmetric backgrounds.

\subsection{$\frac{1}{2}$-BPS Solutions}
\label{sub:1/2BPS}
We now turn to backgrounds that admit only two supercharges. Since we
are interested in solving the Killing spinor equations (\ref{eq:KSE1}) and (\ref{eq:KSE2})
in a 2-dimensional subspace ${\cal S}$ of the full spinor space spanned
by $(\epsilon_{+},\epsilon_{-})$, the integrability condition (\ref{eq:nr_integrability}) will
have to be rederived in the appropriate subspace. For each such space
${\cal S}_{i}$, we will be able to give the necessary and sufficient condition for
integrability. If we label the set of backgrounds that satisfy this
condition by ${\cal M}_{i},$ the full set of $\frac{1}{2}$-BPS backgrounds
is given by $\bigcup{\cal M}_{i}$.

However, since there are of course infinitely many subspaces ${\cal S}$,
using integrability to find all $\frac{1}{2}$-BPS backgrounds seems
impractical. We therefore content ourselves with studying specific
examples of $\frac{1}{2}$-BPS solutions by specifying the subspace
${\cal S}_{i}$ in which their Killing spinors live, and study integrability
for each of them individually.

\subsubsection{\label{sub:2sc}Backgrounds with Two Supercharges of the Form $(0,\epsilon_{-})$}

\label{sub:1/2BPS_epzer}We start by considering the case $\epsilon_{+}=0$.
The Killing spinor equations (\ref{eq:KSE1}) and (\ref{eq:KSE2})
simplify to a single equation:
\begin{align}
\left(D_{\mu}-3S\tau_{\mu}\gamma_{0}\right)\epsilon_{-} & =0.\label{eq:KSE_eplus0}
\end{align}
This equation is integrable if and only if
\begin{align}
R_{\mu\nu}^{\phantom{\mu\nu}ab}(J) & =12\epsilon_{ab}\tau_{[\mu}\partial_{\nu]}S.\label{eq:RJ_eplus0}
\end{align}
In adapted coordinates $R_{ij}^{\phantom{\mu\nu}ab}(J)=0$, which
implies $R_{ij}=0$ (see (\ref{eq:Riem_Gamma})). Thus we can again choose Galilean coordinates
such that $h_{ij}=\delta_{ij}$. The spatial vielbein and its inverse
are given by (\ref{eq:vielb_flat1}) and (\ref{eq:vielb_flat2}), respectively.
The nonzero components of the Riemann tensor are
\begin{align}
R_{\phantom{(i}0j)0}^{(i}(\Gamma) & =-R_{(j0}^{\phantom{(j0}i)}(G)+6\epsilon_{b(i}v^{b}\partial_{j)}S,\label{eq:Riem2SC_1}\\
R_{\phantom{i}0jk}^{i}(\Gamma) & =12\epsilon_{\phantom{i}[k}^{i}\partial_{j]}S,\label{eq:Riem2SC_2}\\
R_{\phantom{i}jk0}^{i}(\Gamma) & =6\epsilon_{\phantom{i}j}^{i}\partial_{k}S.\label{eq:Riem2SC_3}
\end{align}
We can once again express the left hand side of these constraints
in terms of the connection coefficients $\Phi_{i}=\Gamma_{\phantom{i}00}^{i}$,
$C=\frac{1}{2}\epsilon_{ij}\Gamma_{\phantom{i}0j}^{i}.$ Since $R_{(j0}^{\phantom{(j0}i)}(G)$
remains undetermined, the first equation does not impose any further
constraints on $\Phi_{i}$ and $C$. Equations (\ref{eq:Riem2SC_2})
and (\ref{eq:Riem2SC_3}) are equivalent to the condition 
\begin{align}
\partial_{i}S & =\frac{1}{6}\partial_{i}C.\label{eq:SfromC_2SC}
\end{align}
To summarize, a given background admits two supercharges of the form
$(0,\epsilon_{-})$ if and only if $R_{ij}=0$. Given a background
with arbitrary $\Phi_{i}$ and $C$, one can always choose the auxiliary
scalar $S$ such that (\ref{eq:SfromC_2SC}) is satisfied.

The Killing spinors in this class of backgrounds can be constructed
explicitly by solving (\ref{eq:KSE_eplus0}), which now takes the
form
\begin{align}
0 & =\partial_{i}\epsilon_{-},\label{eq:KSE_2SC_1}\\
0 & =\left[\partial_{0}+(\ft12C-3S)\gamma_{0}\right]\epsilon_{-}.\label{eq:KSE_2SC_2}
\end{align}
The second equation can be easily integrated to find the two solutions
\begin{equation}
\epsilon_{-}=e^{\int dt^{\prime}(3S-\ft12C)\gamma_{0}}\epsilon_{0},\qquad(\epsilon_{+}=0),
\end{equation}
where $\epsilon_{0}$ is a constant Majorana spinor. The integrability
condition (\ref{eq:SfromC_2SC}) then guarantees that (\ref{eq:KSE_2SC_1})
is satisfied as well.

\subsubsection{Backgrounds with Two Supercharges of the Form $(\epsilon_{+},0)$ }

\label{sub:1/2BPS_eminzero}Another class of $\frac{1}{2}$-BPS backgrounds
is characterized by $\epsilon_{-}=0$. The Killing spinor equations
in this case read
\begin{align}
\left(D_{\mu}+S\tau_{\mu}\gamma_{0}\right)\epsilon_{+} & =0,\label{eq:KSE_emin0_1}\\
\left(\ft12\omega_{\mu}^{a}\gamma_{a0}-Se_{\mu}^{a}\gamma_{a}\right)\epsilon_{+} & =0.\label{eq:KSE_emin0_2}
\end{align}
Note that the second equation is purely algebraic. Integrability requires
\begin{align}
A_{\mu\nu} & \equiv-\ft14R_{\mu\nu}^{\phantom{\mu\nu}ab}(J)\epsilon_{ab}-2\tau_{[\mu}\partial_{\nu]}S=0,\label{eq:int_emin0_A}
\end{align}
as well as 
\begin{equation}
\ft12\omega_{\mu}^{a}\gamma_{a0}-Se_{\mu}^{a}\gamma_{a}=0.\label{eq:int_emin0_C}
\end{equation}
The first condition implies $R_{\mu\nu}^{\phantom{\mu\nu}ab}(J)=-4\epsilon_{ab}\tau_{[\mu}\partial_{\nu]}S$.
This condition differs from (\ref{eq:RJ_eplus0}) only by a numerical
factor, so we again conclude that $R_{ij}=0$, and choose $h_{ij}=\delta_{ij}$.
In adapted coordinates, the nonzero components of the Riemann tensor
are then
\begin{align}
R_{\phantom{(i}0j)0}^{(i} & =-R_{(j0}^{\phantom{(j0}i)}(G)-2\epsilon_{b(i}v^{b}\partial_{j)}S,\label{eq:Riem_emin0_1}\\
R_{\phantom{i}0jk}^{i} & =-4\epsilon_{\phantom{i}[k}^i\partial_{j]}S,\\
R_{\phantom{i}jk0}^{i} & =-2\epsilon_{\phantom{i}j}^i\partial_{k}S.\label{eq:Riem_emin0_3}
\end{align}
As before, the first equation does not yield any additional constraints.
The second and third equation are equivalent to 
\begin{equation}
\partial_{i}S=-\ft12\partial_{i}C,\label{eq:SfromC_2SC_2}
\end{equation}
which is the analog of (\ref{eq:SfromC_2SC}). 

We now turn to solving the second integrability condition, (\ref{eq:int_emin0_C}).
After evaluating the boost connection $\omega_{\mu}^{a}$ for a metric
of the form
\begin{equation}
h_{\mu\nu}=\left(\begin{array}{cc}
v^{i}v_{i} & -v^{i}\\
-v^{i} & \delta_{ij}
\end{array}\right)
\end{equation}
we arrive at the following conditions:
\begin{align}
S & =-\ft12C+\ft14\epsilon^{ab}\partial_{a}v_{b},\\
\partial_{(a}v_{b)} & =0,\\
\Phi_{a} & =(2S-C)\epsilon_{ab}v^{b}-\partial_0{v}_{b}.
\end{align}
Note that the first two conditions together imply (\ref{eq:SfromC_2SC_2}).
The last condition can be rewritten using (\ref{eq:phiAtomh}) and (\ref{eq:phiC_potentials})
to find
\begin{equation}
\partial_{a}m_{0}-\partial_0{m}_{a}=-\epsilon_{ab}v^{b}\epsilon^{cd}\partial_{c}m_{d}.\label{eq:mv_diff}
\end{equation}
Assume that we fix a background metric $h_{\mu\nu}$ by fixing $v^{a}$,
such that $\partial_{(a}v_{b)}=0$. Then (\ref{eq:mv_diff}) can be
viewed as a constraint on the allowed $m_{\mu}$, which determine
the choice of connection $\Gamma$ in this background. 

To find the two supercharges explicitly, we consider (\ref{eq:KSE_emin0_1}) and (\ref{eq:KSE_emin0_2}):
\begin{align}
0 & =\partial_{i}\epsilon_{+},\\
0 & =\left[\partial_{0}+\ft14\epsilon^{ab}\partial_{a}v_{b}\gamma_{0}\right]\epsilon_{+}.
\end{align}
The solutions are given by 
\begin{equation}
\epsilon_{+}=e^{-\ft14\int dt^{\prime}\epsilon^{ab}\partial_{a}v_{b}\gamma_{0}}\epsilon_{0},\qquad(\epsilon_{-}=0),
\end{equation}
with $\epsilon_{0}$ a constant Majorana spinor.

\subsubsection{Backgrounds with Two Supercharges of the Form $(\epsilon_{+},F\epsilon_{+})$ }
\label{sub:1/2BPS_F}
To complete our discussion of $\frac{1}{2}$-BPS solutions, we consider
the case where the 2-dimensional spinor subspace ${\cal S}$ is not
simply given by $\epsilon_{\pm}=0$, but is rather spanned by nontrivial
linear combinations of $\epsilon_{+}$ and $\epsilon_{-}$. We make the ansatz
\begin{equation}
\epsilon_{-}=F(t,\vec{x})\epsilon_{+}=F^{\mu}(t,\vec{x})\gamma_{\mu}\epsilon_{+}.\label{eq:subspace_F}
\end{equation}
Plugging this ansatz into the Killing spinor equations (\ref{eq:KSE1})
and (\ref{eq:KSE2}), we find
\begin{align}
D_{\mu}\epsilon_{+} & =-S\tau_{\mu}\gamma_{0}\epsilon_{+},\label{eq:F_KSE1}\\
b_{\mu} & =-(\partial_{\mu}+a_{\mu})F^{a}\gamma_{a},\label{eq:F_KSE2}\\
F^{0} & =0.
\end{align}
where 
\begin{align}
a_{\mu} & =-\ft12\omega_{\mu}^{ab}\gamma_{ab}-2S\tau_{\mu}\gamma_{0},\\
b_{\mu} & =\ft12\omega_{\mu}^{a}\gamma_{a0}-Se_{\mu}^{a}\gamma_{a},
\end{align}
encode the geometric information about the background. 

A nonzero Killing spinor $\epsilon_{+}$ exists if and only if (\ref{eq:F_KSE1})
is integrable, which as we saw previously, requires
\begin{equation}
R_{\mu\nu}^{\phantom{\mu\nu}ab}(J)=-4\epsilon_{ab}\tau_{[\mu}\partial_{\nu]}S.
\end{equation}
Following the analysis in section \ref{sub:1/2BPS_eminzero}, the
Riemann components are given by (\ref{eq:Riem_emin0_1})-(\ref{eq:Riem_emin0_3}),
and we again find that $h_{ij}=\delta_{ij}$ and 
\begin{equation}
\partial_{i}S=-\ft12\partial_{i}C.
\end{equation}
It is important to recall that the functions $F^{a}$ were introduced
to determine a certain subspace $(\epsilon_{+},F\epsilon_{+})$ ,
in which we find the Killing spinors. Therefore, (\ref{eq:F_KSE2})
should not be seen as a PDE for $F^{a}$; rather, we should think
of $F$ as being fixed, and (\ref{eq:F_KSE2}) as determining the
background, encoded in $a_{\mu}$ and $b_{\mu}$. In section \ref{sub:1/2BPS_eminzero},
we followed precisely this strategy by choosing $F=0$, which led
to $b_{\mu}=0$. For an arbitrary but fixed $F$, there are two Killing
spinors of the form 
\begin{equation}
\epsilon_{+}=e^{-\int dt(\ft12C+S)\gamma_{0}}\epsilon_{0},\qquad(\epsilon_{-}=F\epsilon_{+}),\label{eq:Kspin_F}
\end{equation}
where $\epsilon_{0}$ is a constant Majorana spinor.

\subsubsection{A Nontrivial Example.}

An example of a nontrivial two-dimensional spinor-subspace ${\cal S}$
with $F\neq0$ is given by 
\begin{equation}
F_{a}=\ft12\epsilon_{ab}v_{b}.
\end{equation}
With this choice, we find the following conditions on the background
fields:
\begin{align}
S & =-\ft12C,\\
\Phi_{a} & =0.
\end{align}
Following (\ref{eq:Kspin_F}), the two supercharges take the form
\begin{equation}
(\epsilon_{+},\epsilon_{-})=(\epsilon_{0},\ft12\epsilon_{ab}v_{b}\gamma^{a}\epsilon_{0}).
\end{equation}
Notice that if the background satisfies $\partial_{\mu}C=0$, all
the conditions for maximal supersymmetry (\ref{eq:max_cond_1})-(\ref{eq:max_cond_3}) are satisfied as well, and supersymmetry is
enhanced from two to four supercharges.

\begin{center}
\begin{figure}
\begin{centering}
\includegraphics[width=12cm]{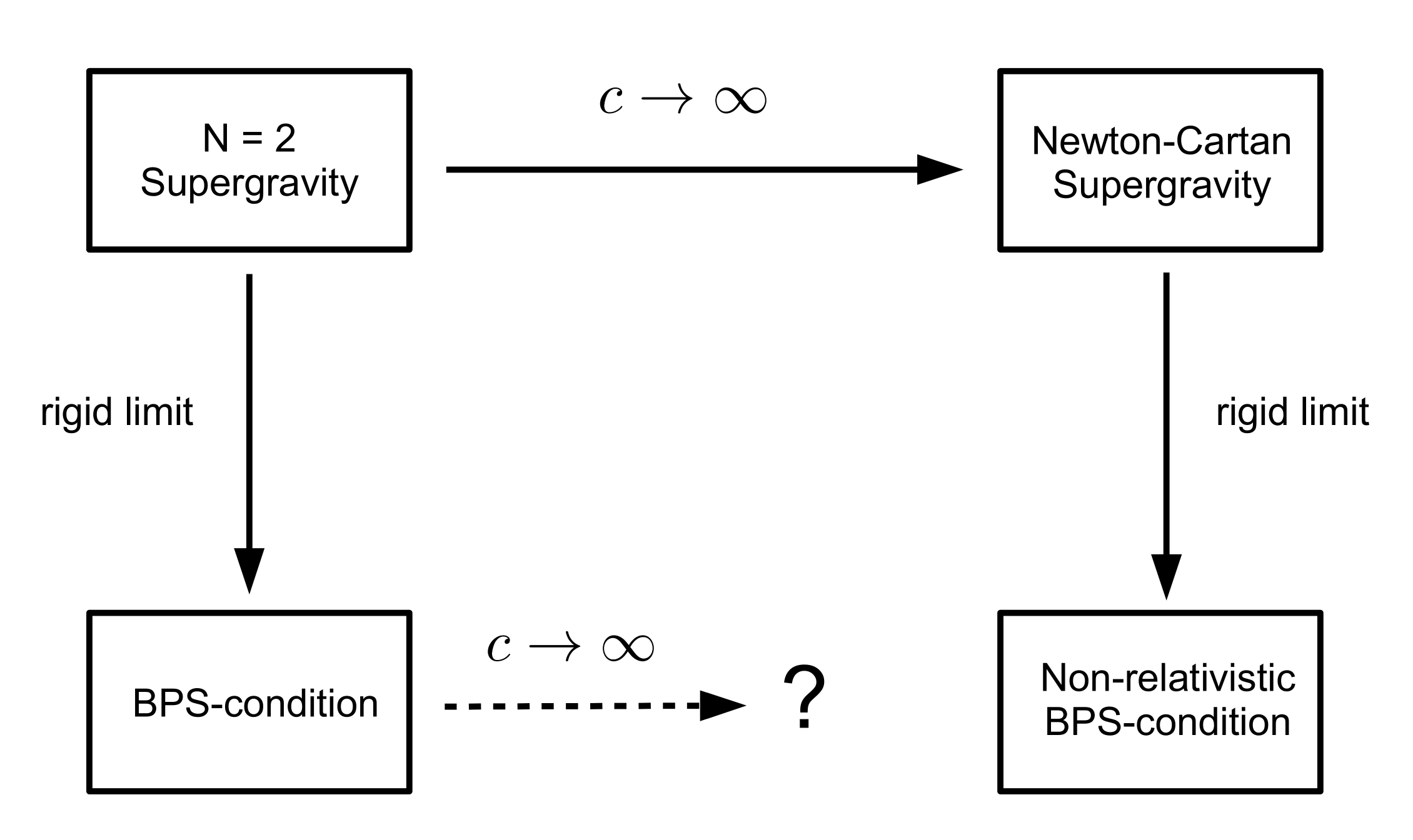}
\caption{\label{fig:diagram_question}}
\par\end{centering}
\end{figure}
\par\end{center}

\section{Rigid Backgrounds of ${\cal N}=2$ Supergravity}

\label{sec:rel_back}
In the previous section we have found rigid backgrounds of non-relativistic supergravity. It is interesting to ask whether the same result can be obtained by taking the non-relativistic limit of rigid backgrounds of $\mathcal{N}=2$ supergravity \cite{Howe:1995zm, Closset:2012ru, Klare:2012gn, Hristov:2013spa, Deger:2013yla} (see figure \ref{fig:diagram_question}).

With this question in mind, in this section we revisit the computation of rigid backgrounds of relativistic supergravity, with the purpose of taking the non-relativistic limit presented in \cite{Bergshoeff:2015uaa} later on. We start by recalling the variation of the gravitino under supersymmetry transformations parametrized by the Majorana spinors $\eta_i$, and an R-transformation parametrized by $\rho$ (see (\ref{eq:gravitinovariation})):
\begin{equation}
\delta \Psi_\mu^i = \mathcal{D}_\mu^{ij} \eta_j -\ft12\epsilon^{ij}\bar{\Psi}_{[\mu i}\Psi_{\nu]j} +\Psi_{\mu j} \rho,
\end{equation}
where the operator $ \mathcal{D}_\mu^{ij}$ given by
\begin{equation}
 \mathcal{D}_\mu^{ij} \equiv \nabla_\mu \delta^{ij} -\gamma_\mu D \delta^{ij} + V_\mu \epsilon^{ij} + \ft14\gamma_\mu \gamma^{\rho\sigma} F_{\rho\sigma} \epsilon^{ij}.
\end{equation}
$F_{\rho \sigma}$ is the field strength of the gauge field $M_\sigma$, $F=dM$. In the rigid limit we set $\Psi_\mu^i=\delta \Psi_\mu^i = 0$, which implies 
\begin{equation}\label{eq:killingspinor}
\mathcal{D}_\mu^{ij} \eta_j =0.
\end{equation}
Rigid supersymmetric backgrounds are given by a choice of $g_{\mu\nu}$, as well as auxiliary fields $V_\mu, M_\mu, D$ such that the Killing spinor equation (\ref{eq:killingspinor}) is integrable.
We determine when solutions to (\ref{eq:killingspinor}) exist by studying its integrability condition, which takes the general form
\begin{equation} \label{eq:integrability}
0=[\mathcal{D}_\mu^{ij},\mathcal{D}_\nu^{jk}] \eta_{k} = (A_{\mu\nu} \delta^{ik} +  B_{\mu\nu}^{\phantom{\mu\nu}\lambda}\gamma_\lambda \delta^{ik} + C_{\mu\nu} \epsilon^{ik} +  D_{\mu\nu}^{\phantom{\mu\nu}\lambda}\gamma_\lambda \epsilon^{ik}) \eta_k.
\end{equation}
In our case, we find
\begin{align}
A_{\mu\nu}& = 0, \label{eq:BPS_A}\\
\label{eq:Bmunulambda}
B_{\mu\nu}^{\phantom{\mu\nu}\lambda} &= -\ft14R_{\mu\nu\rho\sigma} \epsilon^{\rho\sigma\lambda} + 2\delta^\lambda_{\phantom{\lambda}[\mu}\partial_{\nu]}D - 2 \epsilon_{\mu\nu}^{\phantom{\mu\nu}\lambda}D^2 + \ft12\epsilon^{\sigma\tau\lambda}F_{\mu\sigma}F_{\nu\tau},\\
C_{\mu\nu}& = 2\partial_{[\mu} V_{\nu]} + \ft12\epsilon^{\rho\sigma}_{\phantom{\rho\sigma}[\mu}\nabla_{\nu]}F_{\rho\sigma},\\
D_{\mu\nu}^{\phantom{\mu\nu}\lambda} &= \delta^\lambda_{\phantom{\lambda}[\mu}\epsilon_{\nu]}{}^{\rho\sigma}DF_{\rho\sigma} -\ft12\nabla^\lambda F_{\mu\nu}.\label{eq:BPS_D}
\end{align}
We focus on maximally supersymmetric backgrounds, which admit four real supercharges. In this context, just as in the non-relativistic case, there are four linearly independent Killing spinors, so all terms in (\ref{eq:integrability}) should vanish independently. $A_{\mu\nu}=0$ is already guaranteed. $B_{\mu\nu}^{\phantom{\mu\nu}\lambda}=0$ imposes
\begin{align}\label{eq:firstricci}
R_{\mu \nu} &= -8 D^2 g_{\mu \nu} +  g^{\alpha\beta}F_{\alpha\mu}F_{\beta\nu},\\
D&={\textrm{const.}}
\end{align}
These constraints are found by solving for the Riemann tensor and contracting it with the metric, and by contracting Riemann with the Levi-Civita tensor. It is convenient to express the constraints in terms of the dual field strength, defined by
\begin{align} \label{eq:duals}
F_{\mu\nu} &=  \epsilon_{\mu\nu\rho}f^\rho, \nonumber\\
f^\rho &= - \ft12F_{\mu\nu}\epsilon^{\mu\nu\rho}.
\end{align}
With this redefinition, the Ricci tensor (\ref{eq:firstricci}) is then given by
\begin{equation} \label{eq:riccitensor}
R_{\mu\nu} = -8 D^2 g_{\mu\nu} + f_\mu f_\nu - f^\alpha f_\alpha g_{\mu\nu}.
\end{equation}
The condition $C_{\mu\nu}=0 $ imposes a relation between the field strength of $V_\mu$ and $F_{\mu\nu}$,
\begin{equation}\label{eq:ccondition}
 2\partial_{[\mu} V_{\nu]} =- \ft12\epsilon^{\rho\sigma}_{\phantom{\rho\sigma}[\mu}\nabla_{\nu]}F_{\rho\sigma}.
\end{equation}
Using (\ref{eq:duals}), we rewrite (\ref{eq:ccondition}) as an expression relating the field strength of $f_\mu$ and that of $V_\mu$, 
\begin{equation} \label{eq:fandvfluxes}
 2\partial_{[\mu} V_{\nu]} = -\partial_{[\mu}f_{\nu]}.
\end{equation}
Thus, $f_\mu$ and $V_\mu$ are proportional up to the addition of an arbitrary closed 1-form, which would not change (\ref{eq:fandvfluxes}),
\begin{equation} \label{eq:fandv}
f_\mu = - 2 V_\mu + \lambda^\prime_\mu.
\end{equation}
Finally, the condition $D_{\mu\nu}^{\phantom{\mu\nu}\lambda}=0$ implies
\begin{equation} \label{eq:fanddualf}
\nabla_\mu f_{\nu} +2D\epsilon_{\mu\nu\rho}f^{\rho}=0.
\end{equation}
The symmetric and antisymmetric parts of (\ref{eq:fanddualf}) are
\begin{equation} \label{eq:fkilling}
\nabla_{(\mu} f_{\nu)}=0,
\end{equation}
\begin{equation} \label{eq:antifdualf}
\nabla_{[\mu} f_{\nu]} =-2D\epsilon_{\mu\nu\rho}f^{\rho} =-2DF_{\mu\nu} .
\end{equation}
Equation (\ref{eq:fkilling}) shows that $f^\mu$ is a Killing vector, while (\ref{eq:antifdualf}) implies that the field strength of $f_\mu$ is proportional to $F_{\mu\nu}$. Equivalently, $f_\mu$ and $M_\mu$ are related by
\begin{equation} \label{eq:fandm}
f_\mu = -4DM_\mu +\lambda_\mu,
\end{equation}
where $\lambda_\mu$ is an arbitrary closed 1-form. Again, this ambiguity shows up since the addition of a closed 1-form to $f_\mu$ does not change the constraint (\ref{eq:antifdualf}). Another consequence of (\ref{eq:fanddualf}) is
\begin{equation}
\nabla_\mu f^\nu f_\nu = -4D \epsilon_{\mu\nu \rho}f^\nu f^{\rho}  = 0.
\end{equation}
That is, in addition to being a Killing vector, $f^\mu$ has constant norm. Hence, the possible backgrounds are given by $f_\mu=0$ as well as $f_\mu \neq0$ with $f_{\mu}$ timelike, spacelike, or null. 

\subsection{The $f_{\mu}=0$ Case}
\label{sec:fmu0}

For a vanishing $f_\mu$, the Ricci tensor reduces to 
\begin{equation}
R_{\mu\nu} = -8 D^2 g_{\mu\nu}.
\label{eq:einsc}
\end{equation}
The background is locally AdS$_3$, with radius $\ell_A = \frac{1}{2|D|}$. Equations (\ref{eq:fandv}) and (\ref{eq:fandm}) imply that $M_\mu$ and $V_\mu$ are closed and undetermined, i.e., they are pure gauge.

\subsection{The Timelike Case}

For $f^\mu$ a constant norm timelike Killing vector, we introduce adapted coordinates such that
\begin{equation}
f^\mu \partial_\mu = \frac{\partial}{\partial t}.
\end{equation}
Normalizing $f^\mu$ by taking $f^2= -N^2<0$, where $N$ is a non-negative constant, 
we can write the most general metric admitting a timelike Killing direction as 
\begin{equation} \label{eq:timelikemetric}
ds^2 = -N^2(d t + u)^2 + ds^2_{(2)},
\end{equation}
where $u =u_i(x,y)dx^i$ and
\begin{equation}
ds^2_{(2)} =e^{2\sigma}(dx^2 + dy^2)
\end{equation}
is a conformally flat 2-dimensional metric with $\sigma=\sigma(x,y)$. The metric (\ref{eq:timelikemetric}) describes a fibration of a timelike coordinate over the 2-dimensional metric $ds_{(2)}^2$. In the adapted coordinate system, $f_\mu$ is given by
\begin{equation}
f_\mu=(-N^2,-N^2 u_i).
\end{equation}
The integrability constraints (\ref{eq:riccitensor}) and (\ref{eq:fanddualf}) impose 
\begin{equation} \label{eq:timelike2}
e^{-2\sigma}(\partial_1^2 + \partial_2^2)\sigma =(16D^2 - N^2).
\end{equation}
\begin{equation} \label{eq:timelike3}
\frac{4D}{N} = (\partial_1 u_2 - \partial_2 u_1) e^{-2\sigma},
\end{equation}

The first constraint is Liouville's equation, the left hand side of which describes $R_{(2)}$, the Ricci scalar of the 2-dimensional base metric $ds_{(2)}^2$. Liouville's equation has well known solutions; once solved one can insert the solution $\sigma(x,y)$ into the second constraint and solve for $u$, which specifies the way $\mathbb{R}$ is fibered over the spatial manifold. 

We focus on the Ricci scalar, 
\begin{equation} \label{eq:2Dricci}
R_{(2)} = -2e^{-2\sigma}(\partial_1^2 + \partial_2^2)\sigma  = 2(N^2 -16D^2).
\end{equation}
In the last step we used (\ref{eq:timelike2}). We see that the curvature of the 2-dimensional manifold is constant. Note that $N$ contributes positively to the curvature while $D$ contributes negatively. Thus, the supersymmetric backgrounds with a timelike $f_\mu$ are fibrations of a timelike direction over a 2-dimensional manifold that is locally $S^2$, $\mathbb{H}^2$, or $\mathbb{R}^2$:
\begin{align}
&\mathbb{R} \widetilde{\times} S^2, ~\textrm{if}~N>4|D|, \\ 
&\mathbb{R} \widetilde{\times} \mathbb{R}^2, ~\textrm{if}~N=4|D|, \\
&\mathbb{R} \widetilde{\times} \mathbb{H}^2, ~\textrm{if}~N<4|D|.
\end{align}

Up to coordinate transformations, the metric can be obtained by solving (\ref{eq:timelike2}) explicitly:
\begin{equation}
ds^2=-N^2\left(dt+\fft{4D}N\fft{2r^2}{1+(N^2-16D^2)r^2}d\phi\right)^2+\left(\fft2{1+(N^2-16D^2)r^2}\right)^2
(dr^2+r^2d\phi^2).
\end{equation}
For $N>4|D|$, we can transform to spherical coordinates, so that
\begin{align}
ds^2&=L^2\left[-(d\tau-4DL\cos\chi\,d\phi)^2+d\chi^2+\sin^2\chi\,d\phi^2\right],\nn\\
f&=-NL(d\tau-4DL\cos\chi\,d\phi),
\end{align}
where $L^2=(N^2-16D^2)^{-1}$.  When $D=0$, this metric reduces to that of a product space
$\mathbb R\times S^2$.  For $N=4|D|$, the space is flat, and we have
\begin{align}
ds^2&=N^{-2}\left[-(d\tau+\ft12\chi^2d\phi)^2+d\chi^2+\chi^2d\phi^2\right].\nn\\
f&=-(d\tau+\ft12\chi^2d\phi),
\end{align}
Finally, for $N<4|D|$, we obtain instead
\begin{align}
ds^2&=L^2\left[-(d\tau+4DL\cosh\chi\,d\phi)^2+d\chi^2+\sinh^2\chi\,d\phi^2\right],\nn\\
f&=-NL(d\tau+4DL\cosh\chi\,d\phi),
\end{align}
where $L^2=(16D^2-N^2)^{-1}$.  Note that here we cannot obtain a direct product space by setting
$D=0$ because of the strict inequality $|D|>N/4\ge0$.

\subsection{The Spacelike Case}

For $f^2>0$, $f^\mu$ is a spacelike Killing vector; again we can introduce adapted coordinates such that
\begin{equation}
f^\mu \partial_\mu = \frac{\partial}{\partial y}.
\end{equation}
The most general metric admitting a spacelike Killing vector is
\begin{align}  \label{eq:spacelikemetric}
&ds^2 = e^{2\sigma}(-dt^2 + dx^2) + f^2 (dy + u)^2 \\ \nonumber
&\sigma=\sigma(t,x),~u =u_\alpha(t,x)dx^\alpha, \alpha =0, 1.
\end{align}
In the coordinates (\ref{eq:spacelikemetric}) $f_\mu$ is  
\begin{equation}
f_\mu=(f^2u_0,f^2u_1,f^2).
\end{equation}
The metric (\ref{eq:spacelikemetric}) describes a fibration of a spacelike coordinate over a conformally flat Lorentzian manifold. Note that the spacelike and timelike cases can be related via analytic continuation.

The integrability conditions (\ref{eq:riccitensor}) and (\ref{eq:fanddualf}) impose the constraints
\begin{equation}\label{eq:spacelike2}
e^{-2\sigma}(\partial_0^2 - \partial_1^2)\sigma = -(16D^2 + f^2).
\end{equation}
\begin{equation}
-\frac{4D}{f} = (\partial_0 u_1 - \partial_1 u_0) e^{-2\sigma}.
\end{equation}
In complete analogy with the timelike case, the first equation determines the curvature of the 2-dimensional metric $ds_{(2)}^2$ while the second one describes the fibration. The Ricci scalar $R_{(2)}$ is given by
\begin{equation}
R_{(2)} =2 e^{-2\sigma}(\partial_0^2 - \partial_1^2)\sigma = -2(16D^2 + f^2).
\end{equation}
In the last step we used (\ref{eq:spacelike2}). Unlike the timelike case, we see that both $D^2$ and $f^2$ contribute negatively to the curvature. The solution is again a fibration of the real line over a 2-dimensional manifold, but now the only possible 2D manifold is AdS$_2$. Thus, the supersymmetric background allowing for a spacelike Killing vector is $\mathbb{R} \widetilde{\times} \textrm{AdS}_2.$

\subsection{The Null Case}
Finally, we consider the case where $f^{\mu}$ is a null Killing
vector. We define adapted coordinates $(u,v,x)$ such that 
\begin{equation}
f^{\mu}\partial_{\mu}=\frac{\partial}{\partial v}.
\label{eq:adnull}
\end{equation}
Any metric with a null Killing direction $v$ can be written as
\begin{equation}
ds^{2}=H^{-1}\left({\cal F}du^{2}+2dudv\right)+e^{2\sigma}dx^{2},
\end{equation}
where $H$, ${\cal F}$ and $\sigma$ are functions of $u$ and $x$
only. The integrability constraints (\ref{eq:riccitensor}) and (\ref{eq:fanddualf}) translate into the
following differential equations for the metric functions:
\begin{equation}
\partial_{x}\mathrm{log}H=4De^{\sigma},\label{eq:null_diff1}
\end{equation}
\begin{equation}
\partial_{x}^{2}\mathcal{F}-\partial_{x}\mathcal{F}\left(\partial_{x}\mathrm{log}H+\partial_{x}\sigma\right)+2He^{2\sigma}\left[\partial_{u}^{2}\sigma+(\partial_{u}\sigma)^{2}+\partial_{u}\sigma\partial_{u}\mathrm{log}H+H^{-2}\right]=0.\label{eq:null_diff_2}
\end{equation}
This system can be solved by first using (\ref{eq:null_diff1}) to
express $\sigma$ in terms of $H$, plugging the result into (\ref{eq:null_diff_2}),
and then integrating the resulting equation to find ${\cal F}(u,x)$.
However, the solution is cumbersome and not particularly illuminating.
We therefore content ourselves with giving a nontrivial example: consider
the case $H(u,x)=1$, which implies $D=0$ . In this case, $\sigma$ is arbitrary,
and we may choose $\sigma=0$. The solution to (\ref{eq:null_diff_2})
is then given by
\begin{equation}
{\cal F}(u,x)=-x^{2}-a(u)x-b(u),
\end{equation}
with $a(u)$, $b(u)$ being integration constants. The metric reads
\begin{equation}
ds^{2}=2dudv-\left[x^{2}+a(u)x+b(u)\right]du^{2}+dx^{2}.
\end{equation}
This is a plane-fronted wave in Brinkmann coordinates.

\section{Non-relativistic Limit of ${\cal N}=2$ Supergravity}

\label{sec:limit}Given the supersymmetric backgrounds of ${\cal N}=2$
supergravity, it is instructive to study how they connect to the non-relativistic
supersymmetric solutions of section \ref{sec:nonrel_back} in the non-relativistic limit
proposed in \cite{Bergshoeff:2015uaa}. Recall the expansions (\ref{eq:omegaexpansion})
of the background fields: 
\begin{align}
E_{\mu}^{A} & =\delta_{0}^{A}(\omega\tau_{\mu}+\frac{1}{2\omega}m_{\mu})+\delta_{a}^{A}e_{\mu}^{a}, \nonumber\\
M_{\mu} & =\omega\tau_{\mu}-\frac{1}{2\omega}m_{\mu},\label{eq:basicexpansions} \\
D & =\frac{S}{\omega}.\nonumber 
\end{align}
In addition, we will also need the inverse vielbein, which we obtain
perturbatively in $1/\omega$:
\begin{equation}
E_{A}^{\mu}=\delta_{A}^{a}\left(e_{a}^{\mu}-\frac{1}{2\omega^{2}}m_{\nu}e_{a}^{\nu}\tau^{\mu}+{\cal O}(\omega^{-4})\right)+\frac{1}{\omega}\delta_{A}^{0}\left(\tau^{\mu}-\frac{1}{2\omega^{2}}m_{\nu}\tau^{\nu}\tau^{\mu}+{\cal O}(\omega^{-4})\right).
\end{equation}
All other bosonic fields can be expanded in inverse powers of $\omega$.
For example, 
\begin{align}
V_{\mu} & =V_{\mu}^{(0)}+\frac{1}{\omega}V_{\mu}^{(-1)}+\cdots,\nonumber \\
f_{\mu} & =f_{\mu}^{(0)}+\frac{1}{\omega}f_{\mu}^{(-1)}+\cdots.\label{eq:aux_expansion}
\end{align}
In the derivation of the Newton-Cartan supergravity theory \cite{Bergshoeff:2015uaa},
it was necessary to impose the constraints (\ref{eq:constraint_tau}) through (\ref{eq:constraint_F}),
to eliminate divergences. To see if the non-relativistic backgrounds
of section \ref{sec:nonrel_back} could possibly arise as the non-relativistic ($\omega\rightarrow\infty$)
limit of relativistic solutions, we first check if the integrability
conditions (\ref{eq:fandv}) and (\ref{eq:fandm}) are consistent
with (\ref{eq:constraint_V}). Taking $\omega\rightarrow\infty$,
we find: 
\begin{align}
V_{\mu} & =2DM_{\mu}-\ft12(\lambda_{\mu}-\lambda_{\mu}^{\prime})\rightarrow2S\tau_{\mu}-\ft12(\lambda_{\mu}^{(0)}-\lambda_{\mu}^{\prime(0)}).\label{eq:vmu}
\end{align}
We see that relativistic integrability implies a relation between
the auxiliary fields in the non-relativistic limit.
However, there is an ambiguity, parametrized by the closed form $\lambda_{\mu}-\lambda_{\mu}^{\prime}$.
The consistency condition (\ref{eq:constraint_V}) corresponds
to the specific gauge choice $\lambda_{\mu}^{(0)}-\lambda_{\mu}^{\prime(0)}=8S\tau_{\mu}$.

The integrability condition (\ref{eq:riccitensor}), which fixes the
Ricci tensor, can be evaluated in the $\omega\rightarrow\infty$ limit
as well:
\begin{align}
R_{\mu\nu}\rightarrow & ~8S^{2}\tau_{\mu}\tau_{\nu}+\lambda_{\alpha}^{(-1)}\lambda_{\beta}^{(-1)}\eta^{ab}e_{a}^{\alpha}e_{b}^{\beta}\tau_{\mu}\tau_{\nu}.
\end{align}
This expression differs from the non-relativistic integrability condition
(\ref{eq:Ricci_constraint}) only by a $\lambda_{\mu}$-dependent
contribution. The extra contribution can once again be interpreted
as a gauge choice in the definition of the fields, $f_{\mu},M_{\mu},V_{\mu}$:
the particular choice $\lambda_{\mu}^{(-1)}=0$ yields equation (\ref{eq:Ricci_constraint}).

\subsection{Killing Spinor Equation}

The difference between the non-relativistic limit of the ${\cal N=}2$
backgrounds and the non-relativistic solutions found directly within
Newton-Cartan supergravity can be analyzed more systematically by
applying the $\omega\rightarrow\infty$ limit (see section \ref{sec:NC_sugra}) directly
to the relativistic Killing spinor equations (\ref{eq:killingspinor}),
which we recall here for convenience:

\begin{equation}
\nabla_{\mu}\eta^{i}+V_{\mu}\epsilon^{ij}\eta_{j}-\gamma_{\mu}D\eta^{i}+\ft14\gamma_{\mu}\gamma^{\rho\sigma}F_{\rho\sigma}\epsilon^{ij}\eta_{j}=0.\label{eq:killingspinor2}
\end{equation}
Note that the covariant derivative is given by $\nabla_{\mu}=\partial_{\mu}-\ft14\Omega_{\mu}^{AB}\gamma_{AB}$.
We first rewrite (\ref{eq:killingspinor2}) in terms of the spinors
$\eta_{\pm}$ (\ref{eq:plusminusbasis}):
\begin{align}
 & (\partial_{\mu}-\ft14\Omega_{\mu}^{ab}\gamma_{ab})\eta_{+}-\ft12\Omega_{\mu}^{0a}\gamma_{0a}\eta_{-}-D\gamma_{\mu}\eta_{-}-DE_{\mu0}\gamma_{0}(\eta_{-}-\eta_{+})\nonumber \\
 &\kern4em -(V_{\mu}+\ft12f_{\mu})\gamma_{0}\eta_{+}+\ft12F_{\mu\nu}\gamma^{\nu}\gamma_{0}\eta_{-}-\ft12F_{\mu\nu}E_{0}^{\nu}(\eta_{+}+\eta_{-})=0,
\end{align}
\begin{align}
 & (\partial_{\mu}-\ft14\Omega_{\mu}^{ab}\gamma_{ab})\eta_{-}-\ft12\Omega_{\mu}^{0a}\gamma_{0a}\eta_{+}-D\gamma_{\mu}\eta_{+}+DE_{\mu0}\gamma_{0}(\eta_{-}-\eta_{+})\nonumber \\
 &\kern4em +(V_{\mu}+\ft12f_{\mu})\gamma_{0}\eta_{-}-\ft12F_{\mu\nu}\gamma^{\nu}\gamma_{0}\eta_{+}+\ft12F_{\mu\nu}E_{0}^{\nu}(\eta_{+}+\eta_{-})=0.
\end{align}
We can expand these equations in powers of $\omega$ by using the
redefinitions (\ref{eq:spin_conn}) and (\ref{eq:boost_conn}) for the spin/boost-connection
and (\ref{eq:spinorrescaling}) for the Killing spinors, and also
expanding the auxiliary fields according to \eqref{eq:aux_expansion}.
The resulting equations are
\begin{align}
0= & \sqrt{\omega}\bigg[(\partial_{\mu}-\ft14\omega_{\mu}^{ab}\gamma_{ab})\epsilon_{+}-S\tau_{\mu}\gamma_{0}\epsilon_{+}-(V_{\mu}^{(0)}+\ft12f_{\mu}^{(0)})\gamma_{0}\epsilon_{+}+\ft12\tau_{\mu}e_{a}^{\nu}f_{\nu}^{(0)}\epsilon_{-}-\ft12\epsilon_{ab}e_{\mu}^{a}e_{b}^{\nu}f_{\nu}^{(0)}\epsilon_{+}\bigg]\nonumber \\
 & +\frac{1}{\sqrt{\omega}}\left[-(V_{\mu}^{(-1)}+\ft12f_{\mu}^{(-1)})\gamma_{0}\epsilon_{+}+\tau_{\mu}e_{a}^{\nu}f_{\nu}^{(-1)}\gamma^{a}\epsilon_{-}-\ft12\epsilon_{ab}e_{\mu}^{a}e_{b}^{\nu}f_{\nu}^{(-1)}\epsilon_{+}\right]+\mathcal{O}(\omega^{-\frac{3}{2}}),\label{eq:NRexpansion_1}\\
0= & -\frac{\omega^{\frac{3}{2}}}{2}\tau_{\mu}e_{a}^{\nu}f_{\nu}^{(0)}\gamma^{a}\epsilon_{+}-\frac{\sqrt{\omega}}{2}\tau_{\mu}e_{a}^{\nu}f_{\nu}^{(-1)}\gamma^{a}\epsilon_{+}\nonumber \\
 & +\frac{1}{\sqrt{\omega}}\bigg[(\partial_{\mu}-\ft14\omega_{\mu}^{ab}\gamma_{ab})\epsilon_{-}-S\tau_{\mu}\gamma_{0}\epsilon_{-}+(V_{\mu}^{(0)}+\ft12f_{\mu}^{(0)})\gamma_{0}\epsilon_{-}+\ft12\omega_{\mu}^{a}\gamma_{a0}\epsilon_{+}\nonumber \\
 & -Se_{\mu}^{a}\gamma_{a}\epsilon_{+}-\ft12F_{\mu\nu}^{(-1)}e_{a}^{\nu}\gamma^{a}\gamma_{0}\epsilon_{+}\bigg]+\mathcal{O}(\omega^{-\frac{3}{2}}).\label{eq:NRexpansion_2}
\end{align}
Here we have used the definition $F_{\mu\nu}=\epsilon_{\mu\nu\rho}f^{\rho}=E_{\mu}^{A}E_{\nu}^{B}E^{\rho C}\epsilon_{ABC}f_{\rho}$
to expand $F_{\mu\nu}$ in powers of $\omega$ as well. We see that
the Killing spinor equation has split up into terms that are singular/non-singular
in the non-relativistic limit $\omega\rightarrow\infty$. In the full
supergravity approach, the $\mathcal{O}(\sqrt{\omega})$- and $\mathcal{O}(\frac{1}{\sqrt{\omega}})$-terms
would correspond to the variations of $\psi_{\mu+}$ and $\psi_{\mu-}$,
respectively. Here we have already set $\delta\psi_{\mu\pm}=0$ in
the beginning, so \eqref{eq:NRexpansion_1} and \eqref{eq:NRexpansion_2}
lead to Killing spinor equations, plus constraints. Solving \eqref{eq:NRexpansion_1}
and \eqref{eq:NRexpansion_2} order by order in large $\omega$, and
neglecting $\mathcal{O}(\omega^{-\frac{3}{2}})$ terms, we find five
independent equations: there are three constraint equations,

\begin{align}
(V_{\mu}^{(-1)}+\ft12f_{\mu}^{(-1)})\gamma_{0}\epsilon_{+}-\tau_{\mu}e_{a}^{\nu}f_{\nu}^{(-1)}\gamma^{a}\epsilon_{-}+\ft12\epsilon_{ab}e_{\mu}^{a}e_{b}^{\nu}f_{\nu}^{(-1)}\epsilon_{+} & =0,\\
\tau_{\mu}e_{a}^{\nu}f_{\nu}^{(0)}\gamma^{a}\epsilon_{+} & =0,\\
\tau_{\mu}e_{a}^{\nu}f_{\nu}^{(-1)}\gamma^{a}\epsilon_{+} & =0.
\end{align}
Making no further assumptions about the form or number of supercharges,
these conditions need to hold for all $\epsilon_{+}$ and $\epsilon_{-}$.
We thus conclude that
\begin{equation}
V_{\mu}^{(-1)}+\ft12f_{\mu}^{(-1)}=0,\qquad e_{a}^{\nu}f_{\nu}^{(0)}=e_{a}^{\nu}f_{\nu}^{(-1)}=0.
\end{equation}
Using these constraints, we obtain the remaining two equations from
(\ref{eq:NRexpansion_1}) and (\ref{eq:NRexpansion_2}): 
\begin{align}
D_{\mu}\epsilon_{+} & =S\tau_{\mu}\gamma_{0}\epsilon_{+}+\ft12\lambda_{\mu}^{\prime(0)}\gamma_{0}\epsilon_{+},\label{eq:KS_prime1}\\
D_{\mu}\epsilon_{-} & =S\tau_{\mu}\gamma_{0}\epsilon_{-}-\ft12\lambda_{\mu}^{\prime(0)}\gamma_{0}\epsilon_{-}-\ft12\omega_{\mu}^{a}\gamma_{a0}\epsilon_{+}+Se_{\mu}^{a}\gamma_{a}\epsilon_{+}+\ft12\tau_{\mu}(e_{a}^{\nu}f_{\nu}^{(-2)}-\ft12m_{\sigma}e_{a}^{\sigma}\tau^{\nu}f_{\nu}^{(0)})\epsilon_{+}.\label{eq:KS_prime2}
\end{align}
Here $D_{\mu}=\partial_{\mu}-\ft14\omega_{\mu}^{ab}\gamma_{ab}$,
and $\lambda_{\mu}^{\prime}=2V_{\mu}+f_{\mu}$ is the undetermined
closed form introduced in (\ref{eq:fandv}). To obtain the last term in the second
equation, we have further expanded $F_{\mu\nu}$ in powers of $\omega$ using $F_{\mu\nu}=\epsilon_{\mu\nu\rho}f^{\rho}$ as before. 

Comparing the differential equations \eqref{eq:KS_prime1} and \eqref{eq:KS_prime2} with the
non-relativistic Killing spinor equations (\ref{eq:KSE1}) and (\ref{eq:KSE2}),
we see that in general they do not agree. Backgrounds that allow spinor
solutions of \eqref{eq:KS_prime1} and \eqref{eq:KS_prime2} are in general not identical to
the rigid supersymmetric backgrounds we studied in section \ref{sec:nonrel_back}. However,
if we choose

\begin{equation}
e_{a}^{\nu}f_{\nu}^{(-2)}-\ft12m_{\sigma}e_{a}^{\sigma}\tau^{\nu}f_{\nu}^{(0)}  =0,\label{eq:NRlimit_constr1}
\end{equation}
\begin{equation}
\lambda_{\mu}^{\prime(0)}  =-4S\tau_{\mu},\label{eq:NRlimit_constr2}
\end{equation}
we reproduce the non-relativistic Killing spinor equations studied
previously. This means that the backgrounds allowing for solutions
of \eqref{eq:KS_prime1} and \eqref{eq:KS_prime2} are a superset of the maximally supersymmetric solutions of Newton-Cartan
supergravity (see figure \ref{fig:diagram2} in the introduction). 

The difference between the two sets of spinor equations is due the
different order of limits used in their derivation. Recall that to
derive \eqref{eq:KS_prime1} and \eqref{eq:KS_prime2}, we first took the rigid limit $\Psi_{\mu},\delta\Psi_{\mu}\rightarrow0$,
and then the non-relativistic limit $\omega\rightarrow\infty$. On
the other hand, the Newton-Cartan supergravity theory of \cite{Bergshoeff:2015uaa}
was derived by taking $\omega\rightarrow\infty$ first. In this limit,
there are singular terms that arise in the supergravity transformations
with nonzero gravitini. To obtain a consistent theory, these singular
terms have to be eliminated by imposing the following conditions on
the auxiliary fields (see (\ref{eq:constraint_V}) and (\ref{eq:constraint_F})) \cite{Bergshoeff:2015uaa}:
\begin{equation}
\hat{F}_{\mu\nu}=0,\qquad V_{\mu}=-2\tau_{\mu}S.\label{eq:Bergsh_constr}
\end{equation}
In the rigid limit, the first condition becomes $f_\mu=0$. With these constraints, equations \eqref{eq:NRlimit_constr1} and \eqref{eq:NRlimit_constr2}
are satisfied identically, and we obtain the non-relativistic Killing
spinor equations (\ref{eq:KSE1}) and (\ref{eq:KSE2}). Since \eqref{eq:KS_prime1} and \eqref{eq:KS_prime2}
were not derived as a rigid limit of a consistent non-relativistic
supergravity theory, we expect the constraints \eqref{eq:Bergsh_constr}
to reemerge as a consistency condition if one attempts to couple the rigid supersymmetric theory to gravity.

Note, in particular, that the constraint $f_{\mu}=0$ is very strong, as it ought to be imposed before taking the
non-relativistic limit if we wish to remain within the Newton-Cartan supergravity theory of
\cite{Bergshoeff:2015uaa}.  For maximally supersymmetric backgrounds, this restricts the relativistic
starting point to be the $f_\mu=0$ case of section~\ref{sec:fmu0}.  Defining $g_{\mu\nu}=-\omega^2\tau_\mu\tau_\nu
+h_{\mu\nu}$ and substituting into the Einstein condition (\ref{eq:einsc}) then gives
\begin{equation}
R_{\mu\nu}=8S^2\tau_\mu\tau_\nu-8D^2h_{\mu\nu},
\end{equation}
where we have taken $S=\omega D$.  Taking $\omega\to\infty$ along with $D\to0$ while holding $S$ fixed then reproduces
the Ricci condition (\ref{eq:Ricci_constraint}) for maximally supersymmetric solutions of the non-relativistic theory.

\section{Discussion}
In contrast to the maximally supersymmetric case, where we were able to construct all non-relativistic backgrounds explicitly, our discussion of $\frac{1}{2}$-BPS solutions in section \ref{sub:1/2BPS} was limited to providing examples of such backgrounds. To find all backgrounds with reduced supersymmetry, it would be interesting to carry out an analysis using spinor bilinears, to find the necessary and sufficient conditions for preserving a single supercharge (see appendix \ref{sec:bilinear} for such an analysis in the relativistic case).
Nevertheless, the three general cases studied in sections \ref{sub:1/2BPS_epzer}, \ref{sub:1/2BPS_eminzero}, and \ref{sub:1/2BPS_F} essentially capture
all possible $\frac{1}{2}$-BPS solutions. We saw that in each of
these three cases, integrability demands that $R_{ij}=0$, so we can conclude that a necessary condition for non-relativistic
supersymmetry is that spatial slices are flat\footnote{The ansatz in section \ref{sub:1/2BPS_F} can be slightly generalized to $\epsilon_{-}=\left(F^{\mu}(t,\vec{x})\gamma_{\mu}+G(t,\vec{x})\right)\epsilon_{+}$ . However, the integrability condition of (\ref{eq:F_KSE1}) and thus the conclusion $R_{ij}=0$ still remain the same.}. It would be interesting
to see if this continues to be true in higher dimensions, or if it
is possible to allow for nonzero curvature of spatial slices.

In our analysis of non-relativistic $\frac{1}{2}$-BPS solutions,
which admit two supercharges, we may equally have
started by assuming only the form of a single supercharge (e.g. $(\epsilon_{+},0)$).
After solving the integrability conditions in the appropriate subspace of the
4-dimensional space of spinors, we saw that 
Killing spinors necessarily come in pairs, and are characterized by
a two component Majorana spinor $\epsilon_{0}$. Hence a single supercharge
is automatically enhanced to two supercharges, and there are no $\frac{1}{4}$-BPS
solutions. This is a familiar feature from relativistic supersymmetry
\cite{Liu:2012bi, Gauntlett:2002nw} (see also appendix \ref{sec:bilinear}).

In order to make contact with the backgrounds studied in the context of
non-relativistic holography, such as Lifshitz and Schr\"odinger spacetimes,
it is necessary to extend the analysis presented here by including 
nonzero torsion into the supergravity theory.
A torsionful version of Newton-Cartan supergravity has recently been constructed in
\cite{Bergshoeff:2015ija}. It would be interesting to search for rigid supersymmetric
backgrounds within this theory as well, with the goal of systematically
constructing supersymmetric Lifshitz or Schr\"odinger field theories.

With explicit non-relativistic supersymmetric backgrounds now available, the
next step to exploring the concept of non-relativistic supersymmetry
further would be to explicitly construct Lagrangians. Following the ideas
of rigid supersymmetry, one way to accomplish this is to consider
realizations of matter multiplets in Newton-Cartan supergravity and
freeze out gravity to obtain a non-relativistic SUSY algebra. Knowledge of the 
transformation rules then allows one 
to build supersymmetric Lagrangians systematically \cite{Festuccia:2011ws,Knodel:2014xea}.

The study of relativistic supersymmetric field theories has recently
led to a plethora of new results and a deeper understanding
of strongly coupled field theories and holography. Further developing the concepts of non-relativistic supersymmetry and supergravity may turn out to be equally fruitful, and may provide
us with valuable tools to study non-relativistic field theories and
gauge/gravity dualities.

\section*{Acknowledgments}
We would like to thank Anthony Charles, Daniel Mayerson and Leopoldo Pando Zayas for interesting discussions.
This work was supported in part by the US Department of Energy under grant DE-SC0007859.

\appendix

\section{Notation and Conventions}

We choose the 2+1 dimensional Dirac matrices to be
\begin{equation}
\gamma^{A}=\{i\sigma^{2},\sigma^{1},\sigma^{3}\},
\end{equation}
where $\sigma^{i}$ are Pauli matrices, and $A=0,1,2$ denote flat
tangent space indices. The Dirac matrices satisfy the following duality
relations:
\begin{align}
\gamma_{AB} & =-\epsilon_{ABC}\gamma^{C},\\
\gamma_{ABC} & =-\epsilon_{ABC}.
\end{align}
Here $\epsilon_{ABC}$ is the Levi-Civita symbol, with $\epsilon_{012}=1$.
These identities imply the useful relations
\begin{align}
\gamma_{ab} & =\epsilon_{ab}\gamma_{0},\\
\gamma_{a0} & =\epsilon_{ab}\gamma_{b},
\end{align}
where $a,b=1,2$ and $\epsilon_{12}=1$. 

Note that when using curved indices, $\epsilon$ needs to be replaced
by the Levi-Civita tensor $\omega$, so that for example $\gamma_{\mu\nu\rho}=-\omega_{\mu\nu\rho}$.
The Levi-Civita tensor is related to the $\epsilon$-symbol by 
\begin{equation}
\omega_{\mu\nu\rho}=\sqrt{-g}\epsilon_{\mu\nu\rho}.
\end{equation}
We can define a charge conjugation matrix $C=\gamma^{0}$, with the
following properties:
\begin{align}
C^{T} & =C^{-1}=-C=-C^{*}\\
C\gamma^{A}C^{-1} & =-\left(\gamma^{A}\right)^{T}
\end{align}
A Dirac spinor $\psi$ in 2+1 dimensions consists of two complex components.
We define the Dirac conjugate in the usual way as $\bar{\psi}\equiv\psi^{\dagger}\gamma^{0}$,
and the charge conjugate as $\psi^{c}\equiv\psi^{T}C$. Majorana spinors
satisfy the Majorana condition
\begin{equation}
\bar{\psi}=\psi^{c},
\end{equation}
which implies that $\psi$ has two real components.

\section{Bilinear Analysis of ${\cal N}=2$ Backgrounds}
\label{sec:bilinear}

In this appendix, we construct supersymmetric backgrounds preserving at least one supersymmetry
by performing an invariant tensor analysis following the work of \cite{Tod:1983pm,Gauntlett:2002nw}. 
We focus once again on the ${\cal N}=(2,0)$ theory. For a similar analysis in the ${\cal N} = (1,1)$ case, see \cite{Deger:2013yla}.

The $D=3$, $\mathcal N=(2,0)$ superalgebra is specified by a pair of two component
Majorana spinors $\eta^i$ \cite{Howe:1995zm}.  We take $\eta^i$ to be commuting, and form a complete set of
bilinears 
\begin{equation}
\kappa^{[ij]}=\bar\eta^i\eta^j,\qquad K_\mu^{(ij)}=\bar\eta^i\gamma_\mu\eta^j.
\end{equation}
We can equivalently write
\begin{equation}
\kappa=\kappa^{12},\qquad K_\mu=\ft12(K_\mu^{11}+K_\mu^{22}),\qquad
L_\mu^1=K_\mu^{12},\qquad L_\mu^2=\ft12(K_\mu^{11}-K_\mu^{22}).
\label{eq:bilins}
\end{equation}
The set of bilinears comprises one scalar and three vectors, corresponding to ten components,
as expected for the symmetric combination of four spinor components.

The bilinears are not all independent, but may be related via Fierz identities.  The relevant ones are
the norms of the vectors
\begin{equation}
K_\mu K^\mu=-L_\mu^1L^{1\,\mu}=-L_\mu^2L^{2\,\mu}=-\kappa^2,
\end{equation}
the outer product relation
\begin{equation}
L_\mu^1L_\nu^1+L_\mu^2L_\nu^2=K_\mu K_\nu-\eta_{\mu\nu}K_\lambda K^\lambda,
\label{eq:opr}
\end{equation}
and the identities
\begin{equation}
\epsilon_\mu{}^{\nu\rho}L_\nu^1L_\rho^2=\kappa K_\mu,\qquad
\epsilon_\mu{}^{\nu\rho}L_\nu^2K_\rho=-\kappa L_\mu^1,\qquad
\epsilon_\mu{}^{\nu\rho}K_\nu L_\rho^1=-\kappa L_\mu^2,
\label{eq:span}
\end{equation}
demonstrating that the vectors form a basis for the three-dimensional spacetime.

We now turn to the differential identities that may be obtained from the Killing spinor equation
(\ref{eq:gravitinovariation})
\begin{equation}
0=\delta\Psi_{\mu i}=\nabla_{\mu}\eta_{i}+\epsilon^{ij}\eta_{j}V_{\mu}-\gamma_{\mu}\eta_{i}D+\ft14\gamma_{\mu}\gamma_{\rho\sigma}{F}^{\rho\sigma}\epsilon^{ij}\eta_{j}.
\label{eq:kse}
\end{equation}
We find
\begin{eqnarray}
\partial_\mu \kappa^{ij}&=&\ft12\varepsilon_{ij}F_{\mu\nu}K^{kk\,\nu},\\ \nn
\nabla_\mu K_\nu^{ij}&=&2D\epsilon_{\mu\nu\lambda}K^{ij\,\lambda}
-\ft12F_{\mu\nu}\delta^{ij}\varepsilon^{kl}\kappa^{kl}+((\ft14\epsilon_{\mu\lambda\sigma}F^{\lambda\sigma}-V_\mu)\delta_\nu^\rho-\ft12F_{\mu\sigma}\epsilon_\nu{}^{\rho\sigma})(\varepsilon^{ik}K_\rho^{jk}+\varepsilon^{jk}K_\rho^{ik}).\!
\end{eqnarray}
Using (\ref{eq:bilins}), we have
\begin{eqnarray}
\partial_\mu \kappa&=&F_{\mu\nu}K^\nu,\nn\\
\nabla_\mu K_\nu&=&2D\epsilon_{\mu\nu\lambda}K^\lambda-F_{\mu\nu}\kappa,\nn\\
\nabla_\mu L_\nu^a&=&2D\epsilon_{\mu\nu}{}^\lambda L_\lambda^a-\epsilon^{ab}
\epsilon_{\lambda\sigma[\mu}F^{\lambda\sigma}L_{\nu]}^b
-\ft12g_{\mu\nu}\epsilon^{ab}\epsilon^{\rho\lambda\sigma}F_{\lambda\sigma}L_\rho^b
+2\epsilon^{ab}V_\mu L_\nu^b,
\end{eqnarray}
or equivalently
\begin{eqnarray}
d\kappa&=&-i_KF,\nn\\
dK&=&4D*K-2F\kappa,\nn\\
dL^a&=&4D*L^a+2\epsilon^{ab}L^b\wedge*F+2\epsilon^{ab}V\wedge L^b,
\label{eq:diffid}
\end{eqnarray}
along with
\begin{eqnarray}
\nabla_{(\mu}K_{\nu)}&=&0,\nn\\
\nabla_{(\mu}L_{\nu)}^a&=&-\ft12g_{\mu\nu}\epsilon^{ab}\epsilon^{\rho\lambda\sigma}
F_{\lambda\sigma}L_\rho^b +2\epsilon^{ab}V_{(\mu}L_{\nu)}^b.
\label{eq:sdiffid}
\end{eqnarray}
We can immediately see that $K^\mu$ is a Killing vector with norm given by $K_\mu K^\mu=-\kappa^2$.
The analysis then proceeds in two cases: $K^\mu$ being timelike, and $K^\mu$ null.

\subsection{Timelike Case}
If $\kappa\ne0$, $K^\mu$ is a timelike Killing vector.  We proceed by choosing adapted coordinates such that $K^\mu\partial_\mu
=\partial/\partial t$ and writing the metric as
\begin{equation}
ds^2=-\kappa^2(dt+\omega)^2+H^2(dx_1^2+dx_2^2),
\end{equation}
where the metric functions are $\kappa(x^a)$, $\omega_a(x^b)$ and $H(x^a)$.  Introduction of
the natural dreibein basis
\begin{equation}
e^0=\kappa(dt+\omega),\qquad e^a=H\,dx^a,
\end{equation}
allows us to write $K=-\kappa e^0=-\kappa^2(dt+\omega)$.  Acting with the exterior derivative gives
$dK=-2d\kappa\wedge e^0-\kappa^2d\omega$.  Comparison with (\ref{eq:diffid}) then allows us to solve for
$F$
\begin{equation}
F=-e^0\wedge\fft{d\kappa}\kappa-2De^1\wedge e^2+\ft12\kappa d\omega.
\label{eq:Fsol}
\end{equation}
Note that the Bianchi identity $dF=0$ constrains $D(x^a)$ to be independent of $t$.

To proceed, we note that the algebraic identities imply that $L^1$ and $L^2$ span the 2-dimensional
space orthogonal to $e^0$.  Hence we may write
\begin{equation}
L^a=\kappa(\cos\psi\delta^{ab}+\sin\psi\epsilon^{ab}) e^b,
\end{equation}
where $\psi(t,x^a)$ parametrizes a local frame rotation.  Substitution of this expression for
$L^a$ into the identity for $dL^a$ in (\ref{eq:diffid}) allows us to determine $V_\mu$
\begin{equation}
V=\fft12\left(d\psi-(*_2d\omega)e^0+*_2d\log(\kappa/H)\right),
\end{equation}
where $*_2$ is the Hodge dual on the 2-dimensional space spanned by $e^1\wedge e^2$.
There is one remaining condition to check, which is the symmetrized $\nabla_{(\mu}L_{\nu)}^a$
differential identity in (\ref{eq:sdiffid}).  However, explicit computation shows that this is automatically
satisfied for the configuration above.

To summarize, supersymmetric backgrounds with a timelike Killing vector can be written as
\begin{equation}
ds^2=-\kappa^2(dt+\omega)^2+H^2(dx_1^2+dx_2^2),
\end{equation}
along with the auxiliary fields
\begin{eqnarray}
D&=&D(x^a),\nn\\
F=dM&=&-e^0\wedge\fft{d\kappa}\kappa-2De^1\wedge e^2+\ft12\kappa d\omega,\nn\\
V&=&\fft12\left(d\psi-(*_2d\omega)e^0+*_2d\log(\kappa/H)\right).
\label{eq:tcase}
\end{eqnarray}
The solution is specified by the arbitrary (but time-independent) functions
$\kappa(x^a)$, $\omega_a(x^b)$, $H(x^a)$ and $D(x^a)$.  Note that the function $\psi(t,x^a)$ is a gauge parameter, and
can be set to zero if desired.

Given this background field configuration, we can now return to the Killing spinor equation (\ref{eq:kse}).
After some manipulation, we find that the Killing spinors have the form
\begin{equation}
\eta^i=\sqrt{\kappa}(\cos(\psi/2)+\gamma_0\sin(\psi/2))\eta_0^i,
\end{equation}
where $\eta_0^i$ satisfies the $\fft12$-BPS projection
\begin{equation}
\eta_0^i=\gamma_0\varepsilon_{ij}\eta_0^j.
\end{equation}
Although the analysis proceeded by assuming only one unbroken supersymmetry out of four, we see
that the background actually preserves at least two supersymmetries.

While the background (\ref{eq:tcase}) is generically $\fft12$-BPS, the supersymmetry can be
completely unbroken for appropriate choices of the fields.  Such backgrounds ought to match
those obtained by the integrability analysis of Section~\ref{sec:rel_back}.  However, note that
there is no {\it a priori} reason that the choice of metric in (\ref{eq:tcase}) needs to coincide with
the ones of Section~\ref{sec:rel_back}.  In fact, the dual field strength from (\ref{eq:Fsol})
\begin{equation}
f=(2D-\ft12\kappa*_2d\omega)e^0+H^{-1}*_2d\log\kappa,
\end{equation}
does not necessarily even point along a single adapted coordinate direction, and hence falls
outside of the ans\"atze used in Section~\ref{sec:rel_back}.  Of course, we expect the backgrounds
to be related by appropriate coordinate transformations.

\subsection{Null Case}

We now turn to the null case, corresponding to $\kappa=0$.  Following \cite{Gauntlett:2002nw}, we note from
(\ref{eq:diffid}) and (\ref{eq:sdiffid}) that $K^\mu$ satisfies $K\wedge dK=0$ and $K^\mu\nabla_\mu K^\nu=0$,
so it is both hypersurface orthogonal and tangent to affinely parametrized geodesics.  This allows
us to introduce null coordinates $(u,v,x)$ and write
\begin{equation}
K^\mu\fft\partial{\partial x^\mu}=\fft\partial{\partial v},\qquad K_\mu dx^\mu=H^{-1}du.
\label{eq:bilinull}
\end{equation}
We then specialize the metric to take the form
\begin{equation}
ds^2=H^{-1}(\mathcal F\,du^2+2du\,dv)+e^{2\sigma}dx^2,
\end{equation}
where the functions $H(u,x)$, $\mathcal F(u,x)$ and $\sigma(u,x)$ are independent of $v$.  We use the dreibein basis
\begin{equation}
e^{+}=H^{-1}du,\qquad e^{-}=dv+\ft12\mathcal F\,du,\qquad e^3=e^\sigma dx,
\end{equation}
and take the tangent space metric to be $\eta_{+-}=\eta_{33}=1$.

When $\kappa=0$, the first identity in (\ref{eq:diffid}) places a constraint on $F$
\begin{equation}
F=F_{+3}(u,x)e^{+}\wedge e^3,
\end{equation}
where independence of $v$ arises from demanding the Bianchi identity $dF=0$.  The second
identity in (\ref{eq:diffid}) allows us to solve for $D$
\begin{equation}
D=-\ft14e^{-\sigma}\partial_x\log H.
\end{equation}
Note, curiously, that this is similar to the expression (\ref{eq:null_diff1}) obtained from integrability
in the null case, however with the opposite sign.  Nevertheless, there is no inconsistency since the
null Killing vectors (\ref{eq:adnull}) and (\ref{eq:bilinull}) are distinct, so that the corresponding
adapted metrics are not directly equivalent.

Given $F$ and $D$, what remains is to use the differential identities for $L^a$ in (\ref{eq:diffid})
and (\ref{eq:sdiffid}) to solve for $V$.  In order to do so, we note that substituting $\kappa=0$ in (\ref{eq:span})
shows that $L^a\wedge K=0$, so that $L^a$ is parallel to $K$.  This allows us to write $L^a=\phi^aK$
where $\phi^a(u,v,x)$ can in principle depend on all coordinates.  The Fierz identity (\ref{eq:opr}) then
demonstrates that $(\phi^a)^2=1$, so that we can express
\begin{equation}
L^1=K\cos\psi,\qquad L^2=K\sin\psi,
\end{equation}
in terms of a single function $\psi(u,v,x)$.  We can now solve for $V$, and find the simple pure
gauge result
\begin{equation}
V=-\ft12d\psi.
\end{equation}
In summary, for the null Killing vector case, the supersymmetric background is given by
\begin{equation}
ds^2=H^{-1}(\mathcal F\,du^2+2du\,dv)+e^{2\sigma}dx^2,
\end{equation}
along with the auxiliary fields
\begin{eqnarray}
D&=&-\ft14e^{-\sigma}\partial_x\log H,\nn\\
F=dM&=&F_{ux}\,du\wedge dx,\nn\\
V&=&-\ft12d\psi.
\end{eqnarray}
This solution is specified by the metric functions $H(u,x)$, $\mathcal F(u,x)$ and $\sigma(u,x)$ as well
as by $F_{ux}(u,x)$.  As in the timelike case, the function $\psi(u,v,x)$ is a gauge parameter, and can be set
to zero.

Returning to the Killing spinor equation, we find that the Killing spinors have the form
\begin{equation}
\eta=e^{\fft{i}2\psi\sigma^2}\eta_0,\qquad \gamma^1\eta_0=0.
\end{equation}
Here we have used a shorthand notation of combining the two spinor parameters $(\eta^1,\eta^2)$
into a two-component vector which is acted upon by the Pauli matrix $\sigma^2$.  The projection
$\gamma^1\eta_0=0$ demonstrates that this is generically a $\fft12$-BPS background.

\bibliographystyle{JHEP}
\bibliography{NewtonCartan}

\end{document}